\documentclass[twocolumn]{aastex7}
\usepackage{natbib}

\usepackage{subfig}
\usepackage{listings}
\usepackage{amsmath}

\usepackage{xcolor}

\definecolor{codegreen}{rgb}{0,0.6,0}
\definecolor{codegray}{rgb}{0.5,0.5,0.5}
\definecolor{codepurple}{rgb}{0.58,0,0.82}
\definecolor{backcolour}{rgb}{0.95,0.95,0.92}

\lstdefinestyle{mystyle}{
    backgroundcolor=\color{backcolour},   
    commentstyle=\color{codegreen},
    keywordstyle=\color{magenta},
    numberstyle=\tiny\color{codegray},
    stringstyle=\color{codepurple},
    basicstyle=\ttfamily\footnotesize,
    breakatwhitespace=false,         
    breaklines=true,                 
    captionpos=b,                    
    keepspaces=true,                 
    numbers=left,                    
    numbersep=5pt,                  
    showspaces=false,                
    showstringspaces=false,
    showtabs=false,                  
    tabsize=2
}

\begin{document}

\title{The Importance of Standardizing Spectra in the Era of Large Spectroscopic Surveys: \\ 
A Case Study of M Dwarfs in SDSS-V}
\author[0000-0003-3410-5794]{Ilija Medan}
\affiliation{Department of Physics and Astronomy,
	Vanderbilt University,
	Nashville, TN 37235, USA}
\email{ilija.medan@vanderbilt.edu}

\author[0000-0003-0179-9662]{Zachary Way}
\affiliation{Department of Physics and Astronomy, Georgia State University, Atlanta, GA 30302, USA}
\email{zway1@gsu.edu}

\author[0000-0002-0149-1302]{B\'{a}rbara Rojas-Ayala}
\affiliation{Instituto de Alta Investigaci\'on, Universidad de Tarapac\'a, Casilla 7D, Arica, Chile}
\email{brojasayala@academicos.uta.cl}

\author[0000-0003-1479-3059]
{Guy S. Stringfellow}
\affiliation{University of Colorado Boulder, Boulder, Colorado 80309, USA}
\email{Guy.Stringfellow@colorado.edu}

\author[0000-0002-4454-1920]{Conor Sayres}
\affiliation{Department of Astronomy, Box 351580, University of Washington, Seattle, WA 98195, USA}
\email{csayres@uw.edu}

\author[0000-0002-3481-9052]{Keivan G. Stassun}
\affiliation{Department of Physics and Astronomy,
	Vanderbilt University,
	Nashville, TN 37235, USA}
\email{keivan.stassun@vanderbilt.edu}

\author[0000-0003-0174-0564]{Andrew R. Casey}
\affiliation{School of Physics and Astronomy, Monash University, Clayton VIC 3800, Australia}
\affiliation{Center for Computational Astrophysics, Flatiron Institute, 162 Fifth Avenue, New York, NY 10010, USA}
\email{andrew.casey@monash.edu}

\author[0000-0002-2437-2947]{S\'{e}bastien L\'{e}pine}
\affiliation{Department of Physics and Astronomy, Georgia State University, Atlanta, GA 30302, USA}
\email{slepine@gsu.edu}

\author[0000-0002-7512-9453]{Emma Galligan}
\affiliation{Department of Physics and Astronomy, Georgia State University, Atlanta, GA 30302, USA}
\email{egalligan1@gsu.edu}

\author[0000-0002-7883-5425]{Diogo Souto}
\affiliation{Departamento de F\'isica, Universidade Federal de Sergipe, Av. Marcelo Deda Chagas, S/N Cep 49.107-230, S\~ao Crist\'ov\~ao, SE, Brazil}
\email{diogosouto@academico.ufs.br}

\author[orcid=0000-0002-6561-9002,gname=Andrew,sname=Saydjari]{Andrew~K.~Saydjari}
\altaffiliation{Hubble Fellow}
\affiliation{Department of Astrophysical Sciences, Princeton University,
Princeton, NJ 08544 USA}
\email{aksaydjari@gmail.com}

\begin{abstract}

SDSS-V will obtain 100,000s of medium-resolution, optical spectra of M dwarfs with the BOSS instrument. M dwarfs have complex atmospheres, and their spectra contain many wide and dense, overlapping molecular features, so determining accurate stellar parameters by fitting models has been difficult. To circumvent this, other surveys have employed machine learning methods to transfer measurements of stellar parameters from high-resolution spectra to their medium-resolution counterparts. These methods provide large catalogs of stellar parameters but, if not addressed properly, are plagued by biases which are, in part, due to the normalization of the spectra. Typical spectral normalization removes the continuum but preserves the relative depths of the absorption features, but optical M dwarf spectra are almost entirely made up of molecular absorption, which makes this difficult. Here, we develop a \textit{standardization} method that instead defines a pseudo-continuum. We use the spectrum's alpha shape to find the points which lie between the absorption features and apply local polynomial regression to find this pseudo-continuum. To tune the hyperparameters of this method, we create BOSS-like spectra from BT-NextGen models to replicate instrumental, signal-to-noise, and reddening effects. We find that in both this generated set and a validation set of the SDSS-V data, our method performs better than alternative standardizations by producing spectra that are both more uniform for M dwarfs with similar stellar parameters and more easily distinguished compared to M dwarfs of differing parameters. These results from our method will be crucial for better determining stellar parameters of M dwarfs using generative models.


\end{abstract}


\section{Introduction} \label{sec:intro}

There has been a rapid increase in the total number of spectra observed of low-mass stars in recent years due to the onset of large, loosely targeted spectroscopic surveys. This is because low-mass stars, while generally faint, are near-uniformly distributed across the sky, allowing for serendipitous observations whether pointed at the Galactic Center or the Pole. To date, the Fifth Sloan Digital Sky Survey's \citep[SDSS-V;][]{koll2025} BOSS instrument has taken observations of more than 300,000 M dwarf stars with more than 450,000 individual visits. Other spectroscopic surveys that have large numbers of red dwarf spectra include Gaia's XP spectra \citep[$\sim$40 million,][]{gaiaXP}, the LAMOST survey \citep[$\sim$700,000,][]{lamostsdr7}, and DESI \citep[$\sim$560,000][]{DESI}. However, despite the large number of optical spectra of M dwarfs, there remain many outstanding issues in determining their fundamental parameters (e.g. $T_{eff}$, $[M/H]$, and $log(g)$). These issues arise because the M dwarfs lie at the intersection of several compounding complications. They are inherently faint, causing most of their spectra to have low signal-to-noise ratios. They are cool, with many oxide and hydride molecular features dominating their optical spectral energy distribution (SED). Finally, the stellar structure of M dwarfs is dominated by large convection zones \citep{Osterbrock1953}, which have proven particularly difficult to model \citep{Bergemann2017}.
These effects are compounded by reddening and instrument systematics to the point where the optical spectra of the M dwarfs can vary wildly within a particular survey. These issues are not present in higher-mass stars, where the spectra are typically a smooth thermal continuum punctuated by sharp atomic lines. Normalizing these spectra is relatively simple compared to low-mass stars, and the depth of a line relative to the continuum is well-defined.

\begin{figure}
	\centering
	\includegraphics[width=0.5\textwidth]{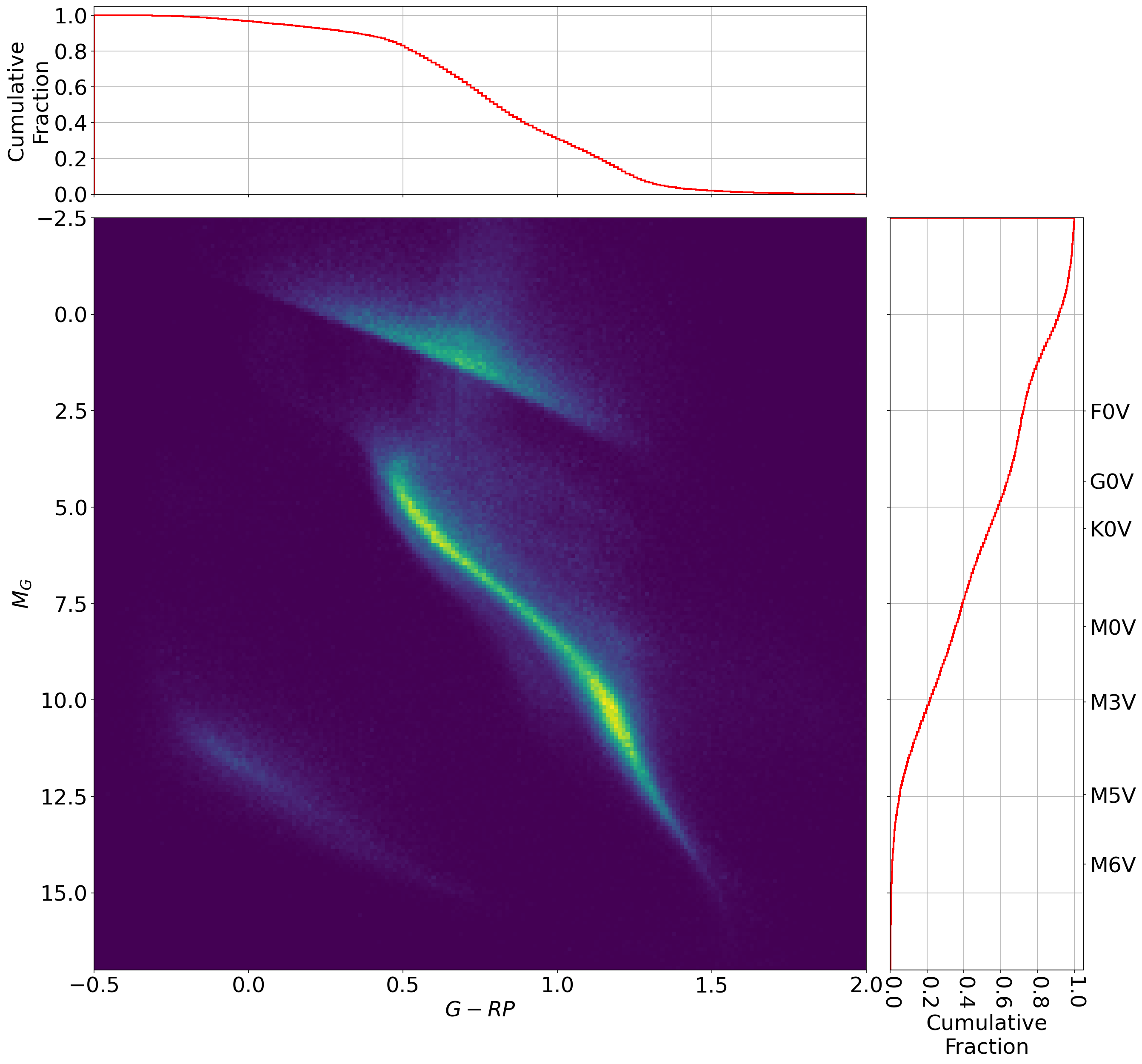}
	\caption{HR diagram of the number of spectra observed with BOSS for MWM during SDSS-V until MJD $= 60715$ (2025/02/09). The top and right histograms show the cumulative distributions of in $G-RP$ and $M_G$, respectively. For the $M_G$ distribution, the location of main sequence spectral types as a function of $M_G$ from \citet{pecaut2013} are shown for reference. From the $M_G$ cumulative distribution on the right, an impressive $\sim$1/3 of BOSS MWM spectra are of M dwarfs.}
	\label{fig:HR_spall}
\end{figure}

However, in the optical portion of an M dwarf's spectrum, there is very little, if \textit{any}, continuum to be found \citep{allard1990, allard1995}. The spectra of M dwarfs are instead dominated by a ``pseudo-continuum", i.e.~regions of relatively constant flux between absorption bands that are not part of the true thermal continuum and vary with the metallicity of the star \citep{Veyette2016}. This strong variation of the pseudo-continuum causes the optical spectra of M dwarfs to be extremely dynamic. The TiO and CaH features are sensitive to the stars' overall metallic abundance and temperature \citep{lepine2007}, resulting in variations in the SED strong enough to affect the broadband photometry (see Fig. \ref{fig:HR_spall}). Because the M dwarfs born at the beginning of the Universe are still alive today, and they are so ubiquitous, they are a crucial element to understanding star and planet formation for the history of the Milky Way. Similarly, M dwarfs can be used as tracers of the local dynamical history, which provides key insight into Galactic evolution.

To fully utilize the optical spectra of M dwarfs, we must account for both the pseudo-continuum and any reduction effects present. For the former, models could be used to define the true continuum of the star and ignore the variations that can arise in the pseudo-continuum. Although there are many model grids built on the assumption of 1D radiative-convective equilibrium conditions available \citep[e.g.,][]{Tsuji1996, allard1997, marcs, kurucz, pheonix}, they all show a wide variation in the shape of the spectrum for a star with the same stellar parameters \citep{sphinx}. Furthermore, no set of models seem to agree with the observational characteristics of M dwarfs. Effective temperatures have been found to be 200--300 K hotter in models compared to observed values \citep{jones2005}. Similarly, radii predictions differ by $\sim$5\% compared to interferometric measurements \citep{charams}. Overall, this demonstrates that it is problematic to rely on M dwarf models to 1) process reduced spectra and 2) to determine fundamental parameters.

Since there are still large discrepancies between the models and the observed optical spectra, a recent trend has been to employ ``label transfer'', particularly for chemical abundances. In label transfer, the stellar parameters from either higher-resolution spectra, where individual lines can be measured accurately \citep{Souto2022}, or from a higher mass star if the M dwarf is in a wide binary system \citep[e.g.,][]{babs2010, newton2014, medan2021, Behmard2025} are used as the ``labels'' in the dataset. Then one can predict the fundamental parameters of other low-mass stars after using these labels for the regression or training of a model. 

These models can either be a generative method (creating a model spectrum from labels) or a discriminative method (predicting labels from the input spectrum). For generative models, normalization is particularly crucial in ensuring the consistency of the results \citep{cannon}. Each method has strengths and weaknesses, but both can result in biases in the final results when applied to M dwarfs. For a complete discussion of the differences between the two models, the introduction of \cite{aspgap} is an excellent resource. Biases in label transfer can arise from the noise in the data itself \citep{ML_bias} but can also arise from the method of data set normalization, as we will show. There has been a wide range of normalization or ``standardization" methods that are incorporated into recent analyses of M dwarf spectra, ranging from completely removing the molecular features \citep{zhang_m_dwarf_spec_TD} to an approximation of \added{the pseudo-}continuum (e.g., GISIC). These methods have been developed and used to analyze spectra from large surveys, but there has not yet been a study of how each standardization affects the recovery of an M dwarf's fundamental parameters.


In this paper, we present two methods to address the issues mentioned above. The first is a method for standardizing the BOSS M dwarf spectra, adapted from \'echelle normalization methods \citep{alpha_shape_1, alpha_shape_2}, where the pseudo-continuum is defined by the alpha shape of the spectrum and local polynomial regression. The second method generates faux, BOSS-like spectra from model spectra, which reflect real flux calibration, reddening, and signal-to-noise variations. With these models, we can show that our Alpha Hulling Standardization can preserve the fundamental parameters of the models, while also reducing the $\chi^2$ of fitting each spectrum. We then compare our method to a few alternatives and find that ours outperforms them in several metrics.

In \S\ref{sec:data} of this paper, we discuss the SDSS-V BOSS spectra of M dwarfs. We define how these stars were targeted, issues in data reduction, and motivate our need for consistent standardization. In \S\ref{sec:method}, we define the Alpha Hulling Standardization Method and introduce the BOSS-like spectra. We use these faux spectra to optimize the hyperparameters of our method, as well as several other ways of standardizing spectra. In \S\ref{sec:results_overall}, we compare how the \added{various methods} perform in recovering the model spectral parameters from the faux spectra and validate that our method also works best on real SDSS-V spectra. In \S\ref{sec:discuss}, we discuss some of the limitations of our method and preview how our Alpha Hulling Standardization can be used with spectra from other surveys, like DESI. We add our concluding remarks in \S\ref{sec:conclusions} and discuss how this method can inform future analysis of SDSS-V and similar optical spectra of M dwarfs.


\section{Data: M Dwarf Spectra in SDSS-V} \label{sec:data}

SDSS-V is an all-sky, multi-epoch spectroscopic survey that will observe over six million objects \citep{koll2025}. It consists of three ``Mappers", which are the top-level programs for the survey. Most relevant here is Milky Way Mapper (MWM), which aims to 1) map the stellar populations and chemodynamics of the Milky Way to understand its evolution and 2) probe stellar physics and system architecture by observing a variety of stars in the Milky Way and Magellanic Clouds. Data for MWM are being collected at two observatories; the Sloan Foundation 2.5-m telescope at Apache Point Observatory (APO) in New Mexico, USA \citep{APO} and the du Pont 100-inch telescope at LCO in Chile \citep{LCO}. Both observatories are equipped with an optical \citep[BOSS;][]{smee13a} and an NIR \citep[APOGEE;][]{wilson19a} fiber-fed spectrograph. Most importantly, SDSS-V has introduced the focal plane system \citep[FPS;][]{FPS}, which utilizes 500 robotic fiber positioners that enable efficient changes between telescope pointings, greatly increasing survey efficiency.

With this increase in survey efficiency, MWM can target a wide range of stars, as described in detail in the DR18 targeting paper \citep{sdssDR18}. Overall, MWM will observe stars across the entire HR diagram, allowing for the study of a wide variety of stellar physics and stellar systems.
With this new hardware and expansive survey strategy for SDSS-V, two things become readily apparent. First, with the FPS, a much larger number of total spectra will be collected over a larger area than in previous iterations of SDSS. Second, with the science objectives of SDSS-V, there are many overlapping areas that will utilize low-mass stars for their science. This is especially true for optical BOSS spectra. While SDSS-V will provide APOGEE spectra for more unique objects, as most of the fibers in the FPS are for BOSS most of the exposures will be of BOSS spectra for MWM. With multiple exposures of the same object, BOSS will also allow us to probe to lower magnitudes. Figure \ref{fig:HR_spall} is an HR diagram of the $\sim2,200,000$ BOSS spectra observed with BOSS for MWM during SDSS-V through MJD $= 60715$, which corresponds to a calendar date of 2025/02/09 (YYYY/MM/DD). This HR diagram and the cumulative distribution of spectra as a function of $M_G$, demonstrate that about 1 in 3 MWM BOSS observations are of M dwarfs. M dwarf spectra will be ubiquitous in the dataset resulting from SDSS-V and this is why it is imperative that methods are developed with M dwarfs as the primary focus.

The M dwarf spectra of interest in this work are reduced using the BOSS pipeline \citep{Bolton2012, Dawson2013,morrison2025}. Here we provide a brief summary of the processing of these spectra for SDSS-V. The pipeline begins with the reduction of the 2D spectra \citep[\texttt{spec2d}; see][]{Stoughton2002}. Here, raw electrons are extracted from CCD images row-by-row using the Gaussian fitting technique from \citet{Horne1986}. Several calibration frames are then used to calibrate these counts. Fiber flats are taken from the illumination of arc lamps to correct for fiber-to-fiber variations. In every observation, several fibers are reserved for sky positions, which allows for a model of the sky to be calculated and interpolated to the location of every science fiber. These two calibrations are then used to get the flat-fielded and sky-subtracted counts. Then, to calibrate the flux of the spectra, the pipeline utilizes both flux correction vectors (low-order polynomial per-fiber fits to make different exposures of the same object consistent) and flux distortion vectors (a model of the variation on the throughput across the focal plane). With these corrections, the 1D flux calibrated spectra are produced. The resulting reduced spectra have $R = 1560$ at 3700 \AA \ to $R= 2270$ at 6000 \AA \ (blue channel), and $R = 1850$ at 6000 \AA \ to $R = 2650$ at 9000 \AA  \ (red channel). The spectra cover the full range of $\sim 3600 - 10,400$ \AA.

\begin{figure}
	\centering
	\includegraphics[width=0.45\textwidth]{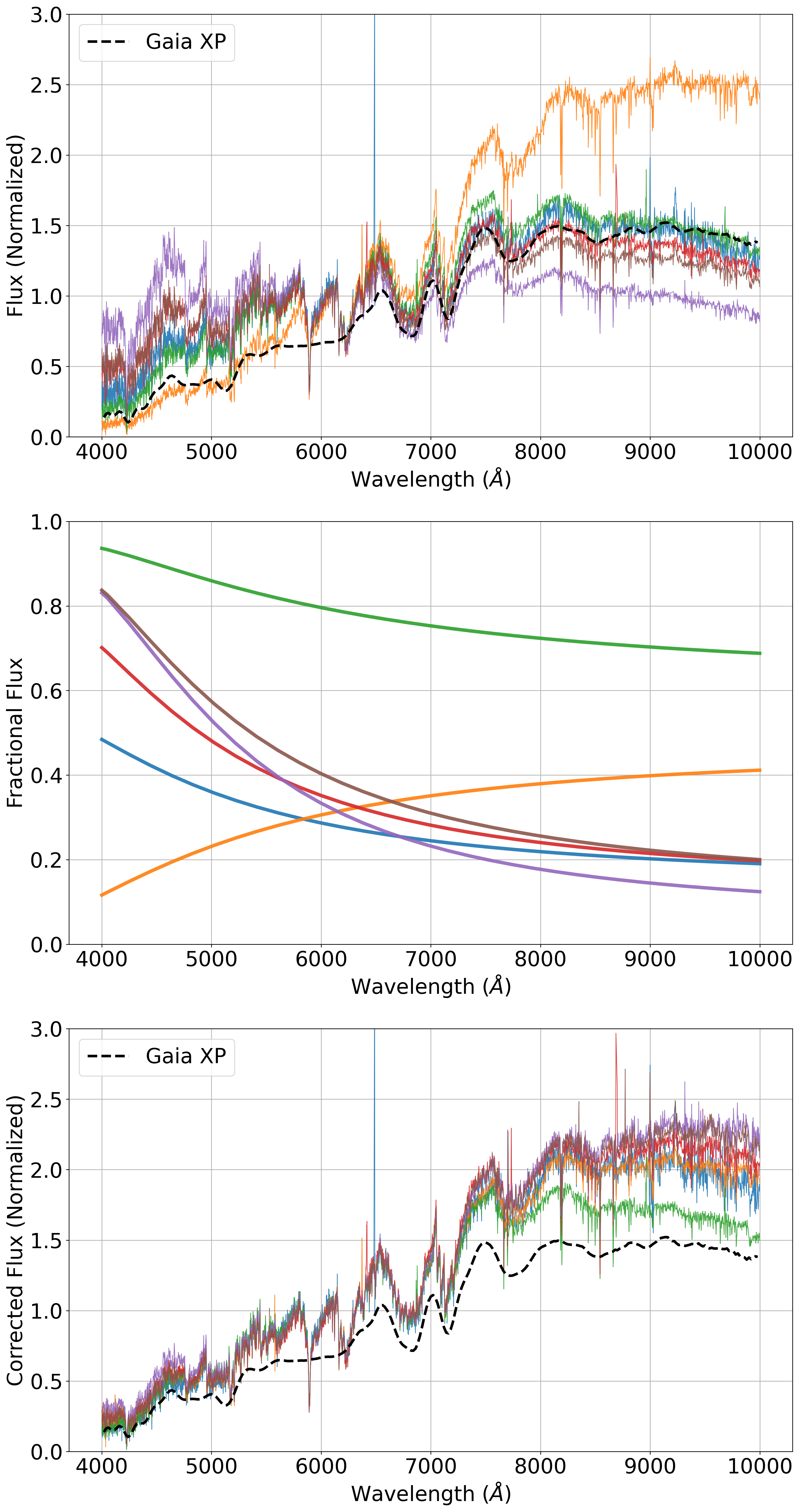}
	\caption{\textbf{\textit{Top Panel:}} The median-normalized BOSS spectra of the same M dwarf observed at different epochs. The BOSS spectra shown were reduced using \texttt{v6\_2\_0} of the BOSS pipeline and include data up until MJD $= 60715$ (2025/02/09). The dashed black line shows the median normalized Gaia XP spectrum of the star, for reference. \textbf{\textit{Middle Panel:}} The estimated fraction of flux entering the fiber for each observation of the M dwarf shown in the top panel. \textbf{\textit{Bottom Panel:}} The median-normalized BOSS spectra from the top panel after dividing by the fractional flux curves in the middle panel. The dashed black line shows the median normalized Gaia XP spectrum of the star, for reference.}
	\label{fig:ex_var_mdwarf}
\end{figure}

Unfortunately, there are major inconsistencies in the reduced spectra. This can be seen by looking \added{at} spectra taken \added{on} different MJDs, but of the same source. The top panel of Figure \ref{fig:ex_var_mdwarf} shows an example of an M dwarf with 6 epochs of data. Not only does the absolute value of the flux change at a given wavelength, but the shape of the spectrum also changes from observation to observation. This is most clearly seen when, for example, the orange spectrum has the largest flux values $>7500$ \AA \ but then has the lowest flux values $<5500$ \AA.

\begin{figure*}
	\centering
    \includegraphics[width=\textwidth]{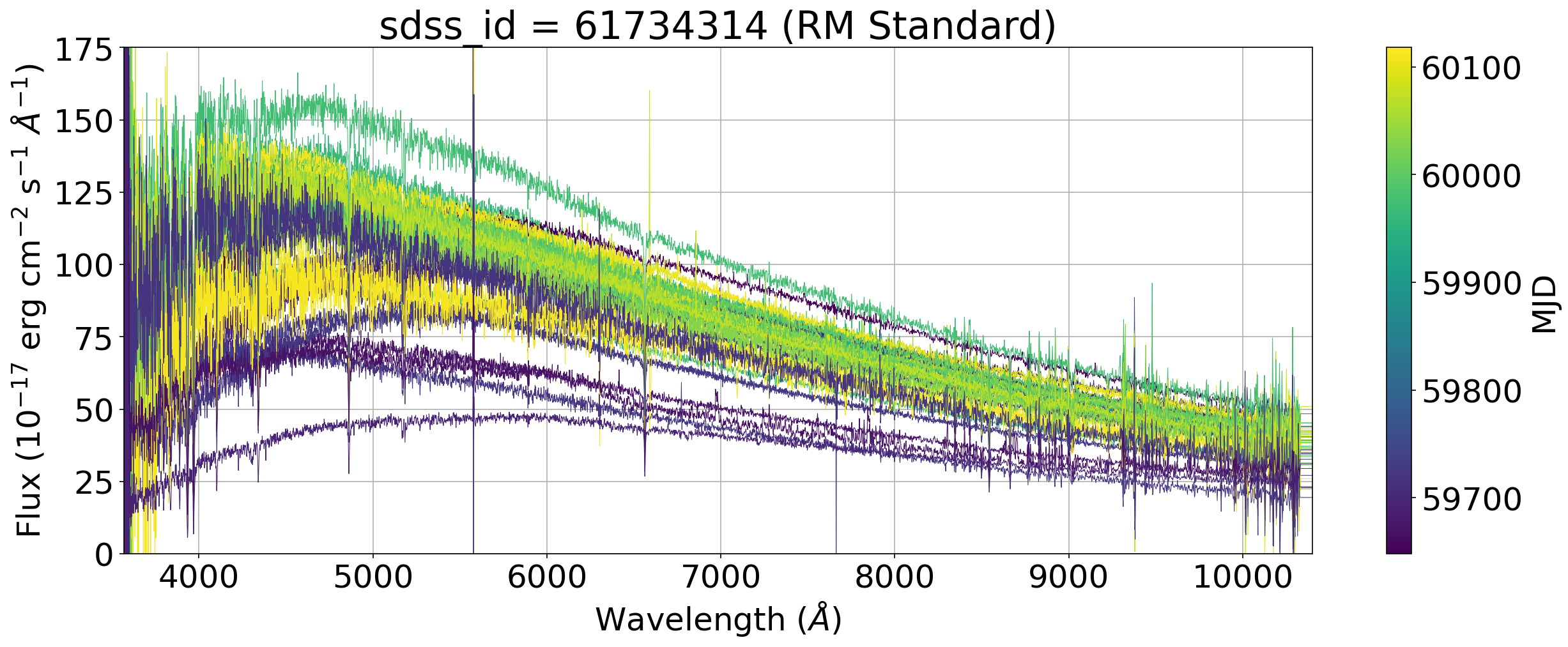}
	\caption{The spectra of the same standard source observed at different MJDs in a Reverberation Mapping (RM) field. This standard star has 54 epochs of data. It is important to note for the RM fields, that the standard is always observed with the same fiber in the field. Significant changes in the spectral shape are observed. The above spectra were reduced using \texttt{v6\_2\_0} of the BOSS pipeline and include data up until MJD $= 60715$ (2025/02/09).}
	\label{fig:ex_var_rm}
\end{figure*}

M dwarfs can of course be very dynamic sources, we assert that the above variations are not likely astrophysical. However, to provide stronger evidence that these variations are nonphysical, we can look at the stars used as standards for the BOSS pipeline. A prime testing ground for this is in the Reverberation Mapping (RM) fields. These fields will be observed for $>100$ epochs, and every design is identical to one another. This means that they include the same standards observed on the same fiber. Figure \ref{fig:ex_var_rm} shows spectra from 54 epochs of the same standard in one of these fields. Like the M dwarf example, we see significant changes in the absolute flux at a given wavelength. More worrying though is the fact that even with this standard, which should have a constant spectral shape, we see large variations in the overall SED as the peak in flux of the spectrum changes for different observation dates.

Both examples in Figures \ref{fig:ex_var_mdwarf} and \ref{fig:ex_var_rm} are some of the more extreme ones, but we do find that such variations exist at some level in most of the BOSS M dwarf spectra. This SED variation could be due to multiple factors but can be largely explained by fiber positioning errors.  To motivate this, we will briefly focus on the six M dwarf BOSS spectra plotted in the top panel of Figure \ref{fig:ex_var_mdwarf}.  This star is a bright object that requires an intentional fiber offset \citep[$\sim1.5$ arcseconds; see more details in][]{fpsdesign_paper} to avoid saturating the BOSS detectors in a 15-minute exposure, so the spectroscopic fiber is collecting flux from only the wings of the PSF.  In this regime, the total flux collected will be sensitive to the seeing conditions on a given night of observation.  The spectra in the top panel of Figure \ref{fig:ex_var_mdwarf} are median normalized to roughly remove a total flux offset component.  We additionally show the median normalized Gaia XP spectrum for this star for reference.  We observe high variance in the overall shape of the SED for this object, which highlights a flux-calibration issue.

The SDSS-V 2.5m telescopes are not equipped with an Atmospheric Dispersion Corrector so the wavelengths of light entering the spectroscopic fiber are subject to the airmass-dependent effects of differential atmospheric refraction (DAR).  Currently, the BOSS pipeline does not consider how DAR in the presence of fiber offsets (whether intentional or accidental) might change the observed SED of a given object.  Here we can estimate the flux response in the fiber as a function of wavelength in an ex-post-facto fashion for these specific observations.

Each time the robots are configured, a Fiber View Camera \citep[FVC;][]{Jurgenson2020} is used to measure and record the positions of the fibers in the focal plane.  While the spectrograph is exposing, we can derive the exact telescope pointing, airmass, and seeing conditions for a given observation from guide camera images.  With these pieces of information, we can estimate the star-to-fiber offset, and thus the amount of light entering the fiber in the presence of DAR.  We show these wavelength-dependent curves for each observation in the middle panel of Figure \ref{fig:ex_var_mdwarf}. A value of 1 in this plot indicates there is no flux attenuation (fiber is perfectly centered on the object for that wavelength), a value of 0.5 means half of the flux is lost due to a relative offset between the star and the fiber.  Finally, the bottom panel of Figure \ref{fig:ex_var_mdwarf} shows the results of first dividing the spectra by the attenuation curves in the middle panel of Figure \ref{fig:ex_var_mdwarf} before median normalization.  Clearly, the SED variance is greatly reduced, although perhaps still biased with respect to the Gaia XP spectrum of the source.  Within the BOSS data teams, there is ongoing work to implement flux calibration work like this for future pipeline versions.



Overall, this demonstrates that it is crucial that issues in flux calibration are accounted for in the current SDSS data releases, particularly for analysis pipelines that rely on matching templates with spectra or for the training of generative models for label prediction. These methods would then require some kind of standardization of the spectra to negate the issues in flux calibration that we see here. In this work, we will focus on the SDSS-V BOSS spectra for M dwarfs and develop a method to standardize the spectra that is specifically tailored for them. The goal is for the final, standardized spectra to be absent of these flux calibration issues, such that they will be better suited for future data analysis pipelines and preserve the underlying, intrinsic parameters of the star.

\section{Methods: Spectral Standardization }\label{sec:method}

\subsection{Alpha Hulling Method}

Here we lay out the Alpha Hulling Standardization Method that we will use to standardize the BOSS M dwarf spectra. The following sections describe the method and how it is optimized for the BOSS M dwarfs.

\subsubsection{Description}\label{sec:method_descript}

The method outlined here is heavily inspired by a normalization method designed for \'echelle spectra, where the continuum is found using the alpha shape of the data \citep{alpha_shape_1, alpha_shape_2}. An alpha shape is a generalization of the convex hull, where the outlining boundary of the data can be made more ``crude" or ``fine" depending on the value of $\alpha$ \citep{alpha_shape_def}. 
In two dimensions, the convex hull of a set of points is simply the smallest convex shape that encloses all the points. The analogy used by \cite{alpha_shape_2} is that if someone stretched a rubber band around the data, the shape of the resulting stretched band would be the convex hull. But convex hulls do not allow for inflection points\added{. So,} for spectra we can use the generalized alpha shapes, which allow for internal angles below $<180^\circ + \alpha$ (rather than strictly $<180^\circ$).



So, how does one determine the alpha shape of a spectrum? It is best to imagine that there is a ball of radius $1/ \alpha$ that we want to move around our spectrum. Whenever the ball intersects any two points in the spectrum but does not contain any other points within the area of the ball, these are added to the perimeter of our alpha shape. 
We can then tune the radius of our ball to capture nearer or further points at the edge of our alpha shape. A small ball will yield a very tight alpha shape around the data, slotting into the molecular features, while a large ball will roll over the top of most of the spectrum. After defining the alpha shape, we can select the points along the surface, which should be the top of our spectrum, fit some line through them and this defines our pseudo-continuum. 

It is clear from the above that depending on how we define the spectrum (e.g., masking pixels, emission features, smoothing of spectrum), the size of our alpha ball and how we fit the top points of our alpha shape, we can greatly change the resulting pseudo-continuum. In particular, a simple polynomial regression may not be flexible enough for the shape of the pseudo-continuum. So, we will instead use local polynomial regression \citep{cleveland1979robust}, as was done in \cite{alpha_shape_1}.

Local polynomial regression is a non-parametric method for fitting a line through a set of points. For each point $x_i$ you want to evaluate, a nth degree polynomial is fit to the data, where the neighboring data are weighted by a kernel function centered on the given point. The value of $x_i$ is the evaluation of that polynomial. This process is then repeated for all evaluation points within the bounds of your data. For a polynomial of $n=0$ degrees, this would be equivalent to a weighted moving average. This method of non-parametric fitting allows you to smoothly fit the data and control how much nearby points influence that smoothing. For local polynomial regression, the main parameters to tune are the degree of the polynomial, the function of the kernel, and the size of said kernel.

\begin{figure}
	\centering
	\includegraphics[width=0.31\textwidth]{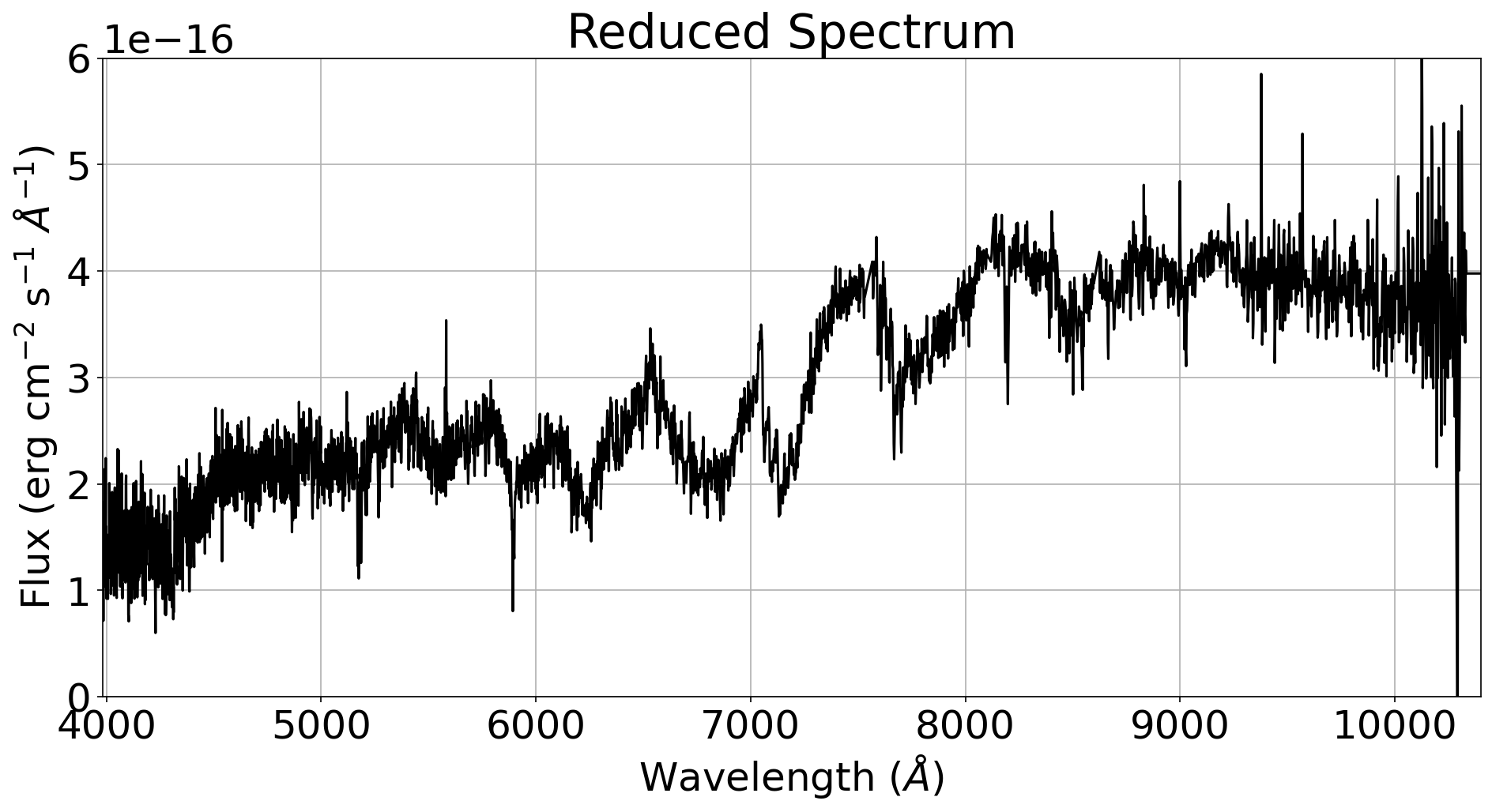}
    \includegraphics[width=0.31\textwidth]{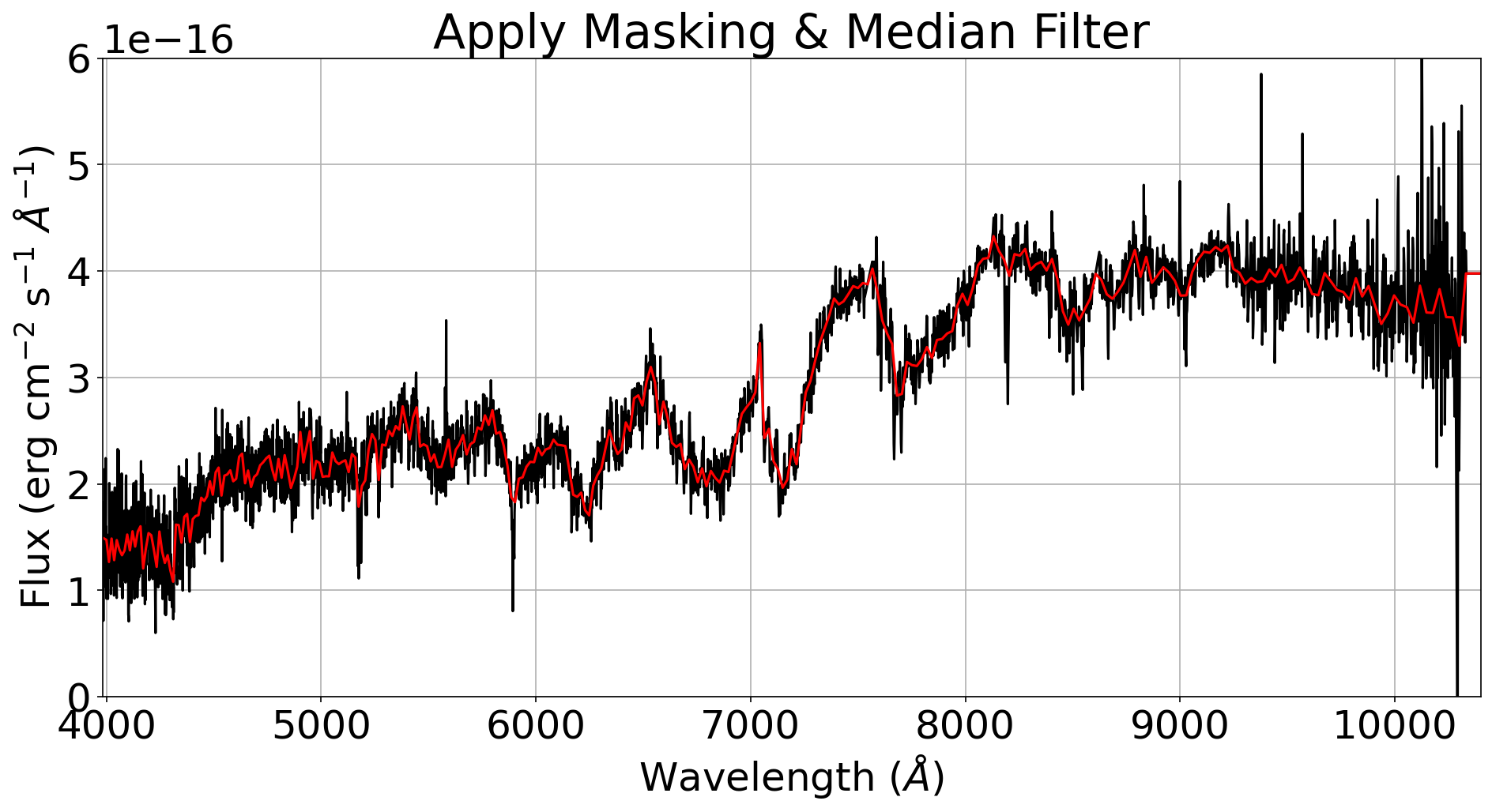}
    \includegraphics[width=0.33\textwidth]{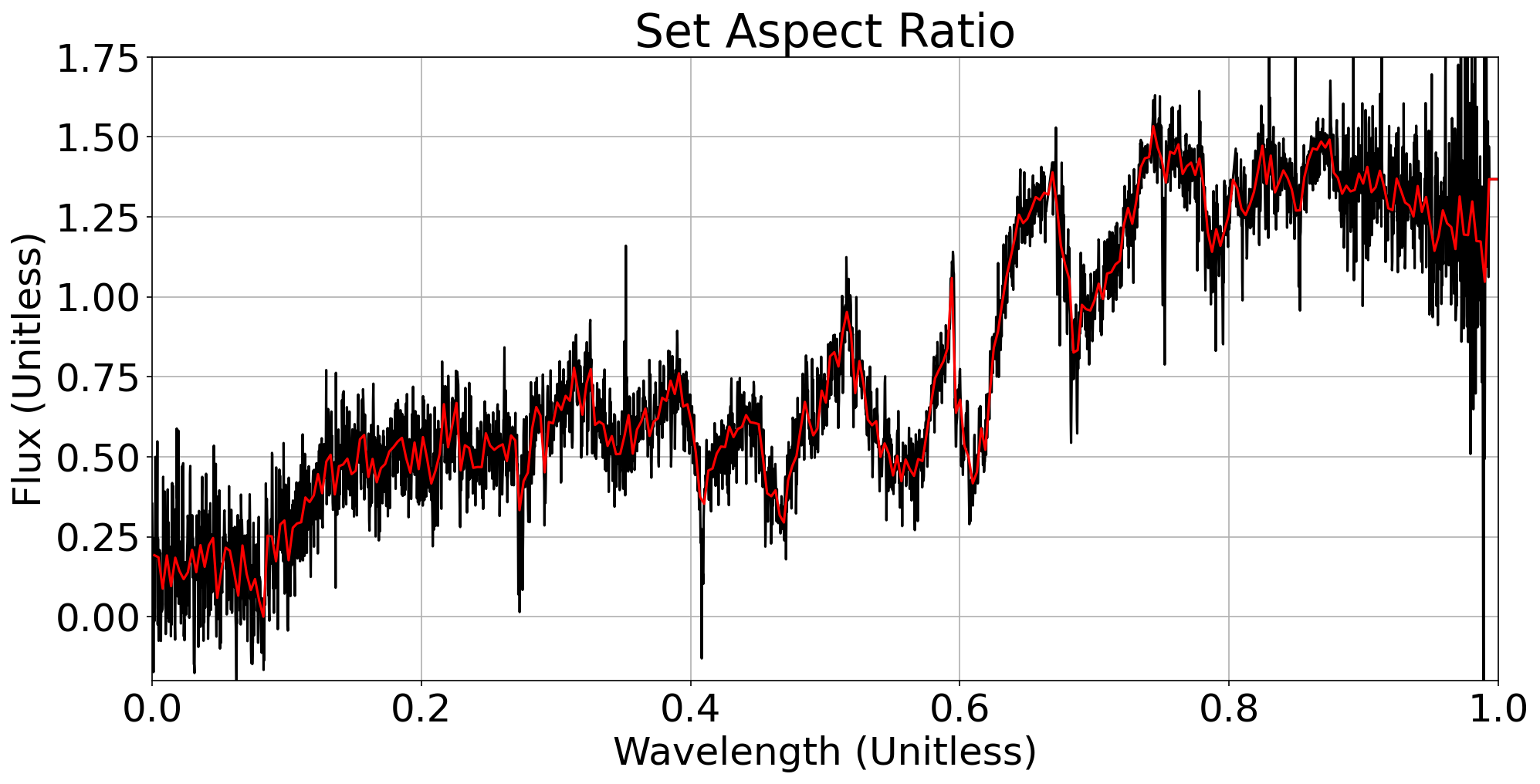}
    \includegraphics[width=0.33\textwidth]{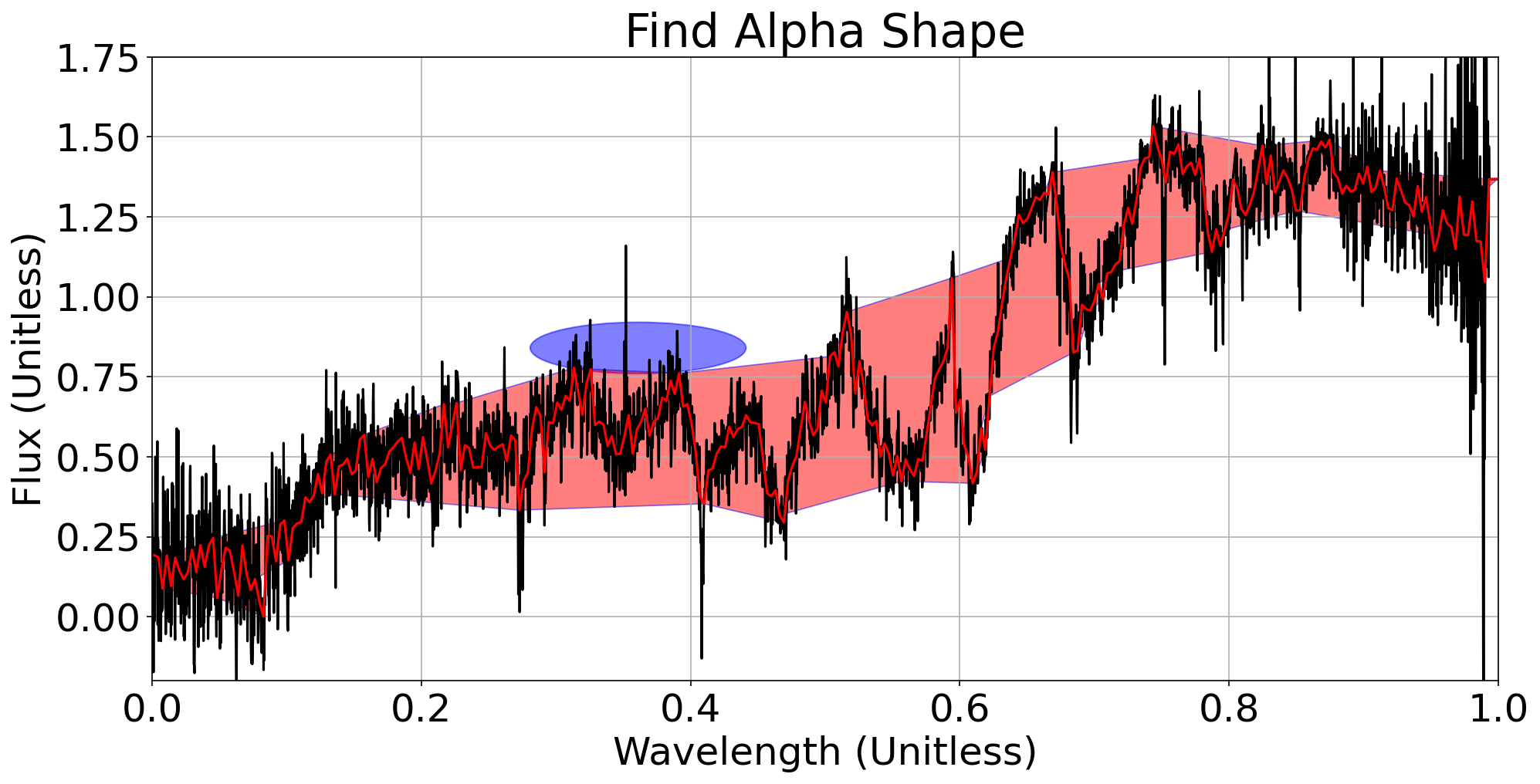}
    \includegraphics[width=0.33\textwidth]{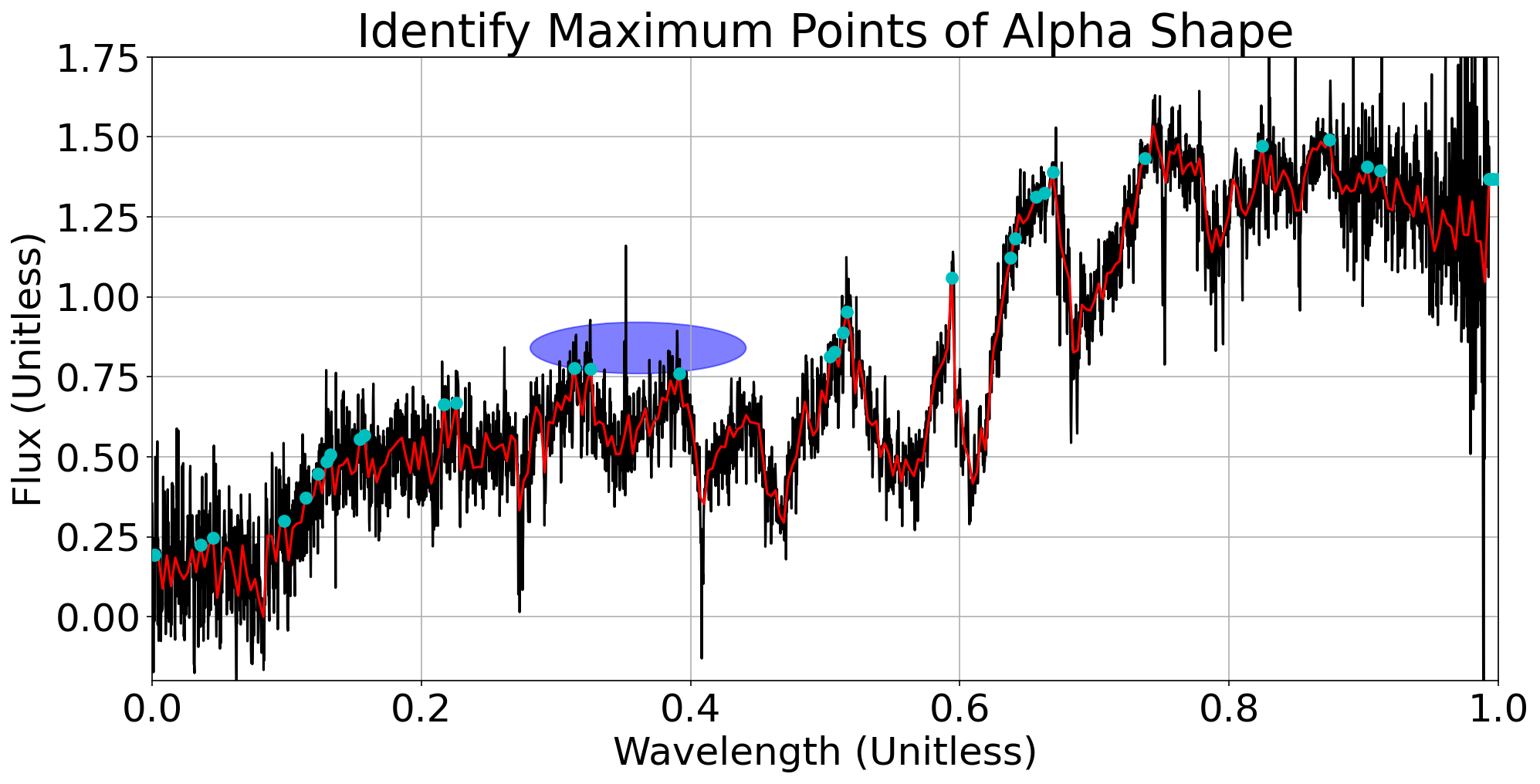}
    \includegraphics[width=0.33\textwidth]{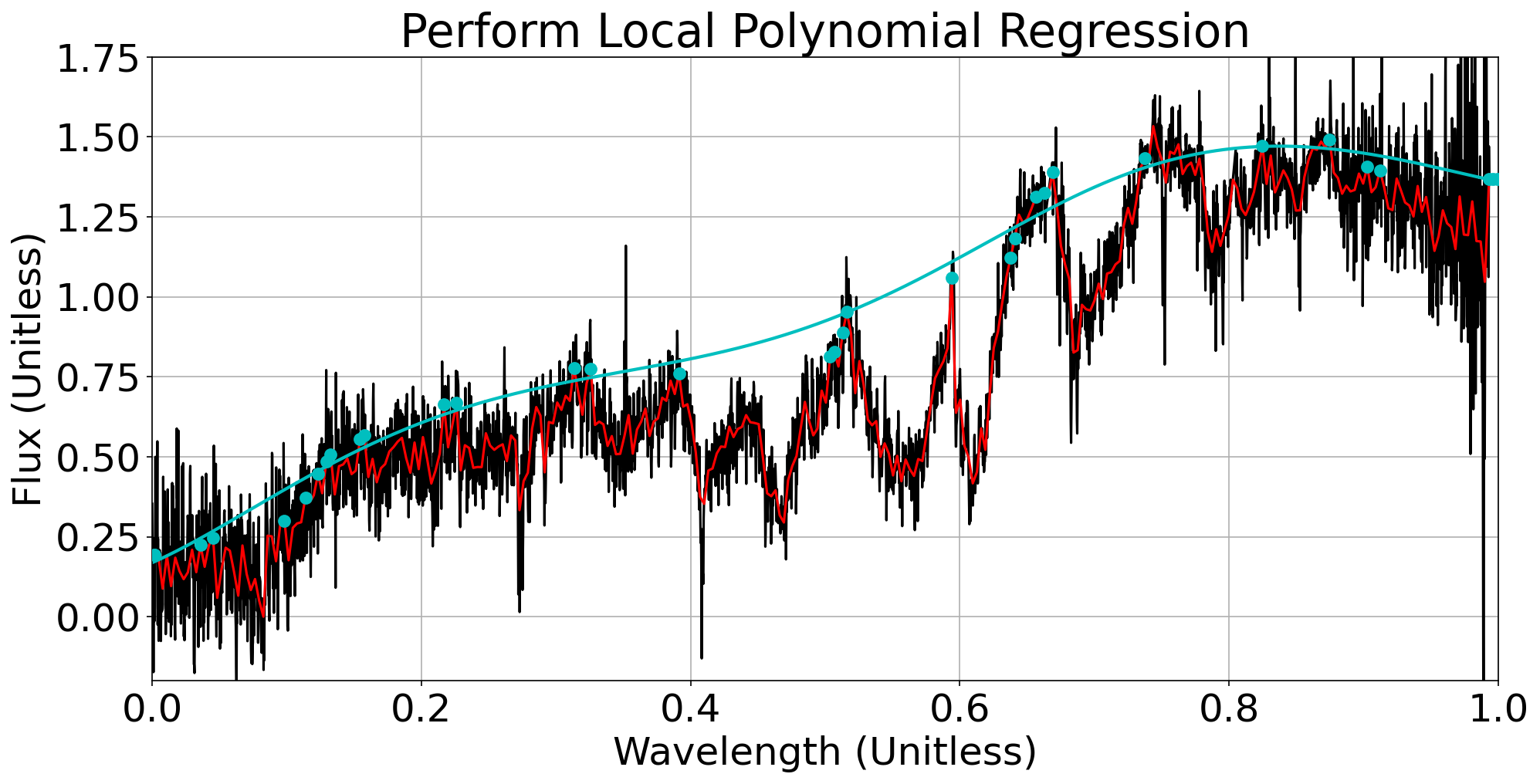}
    \includegraphics[width=0.33\textwidth]{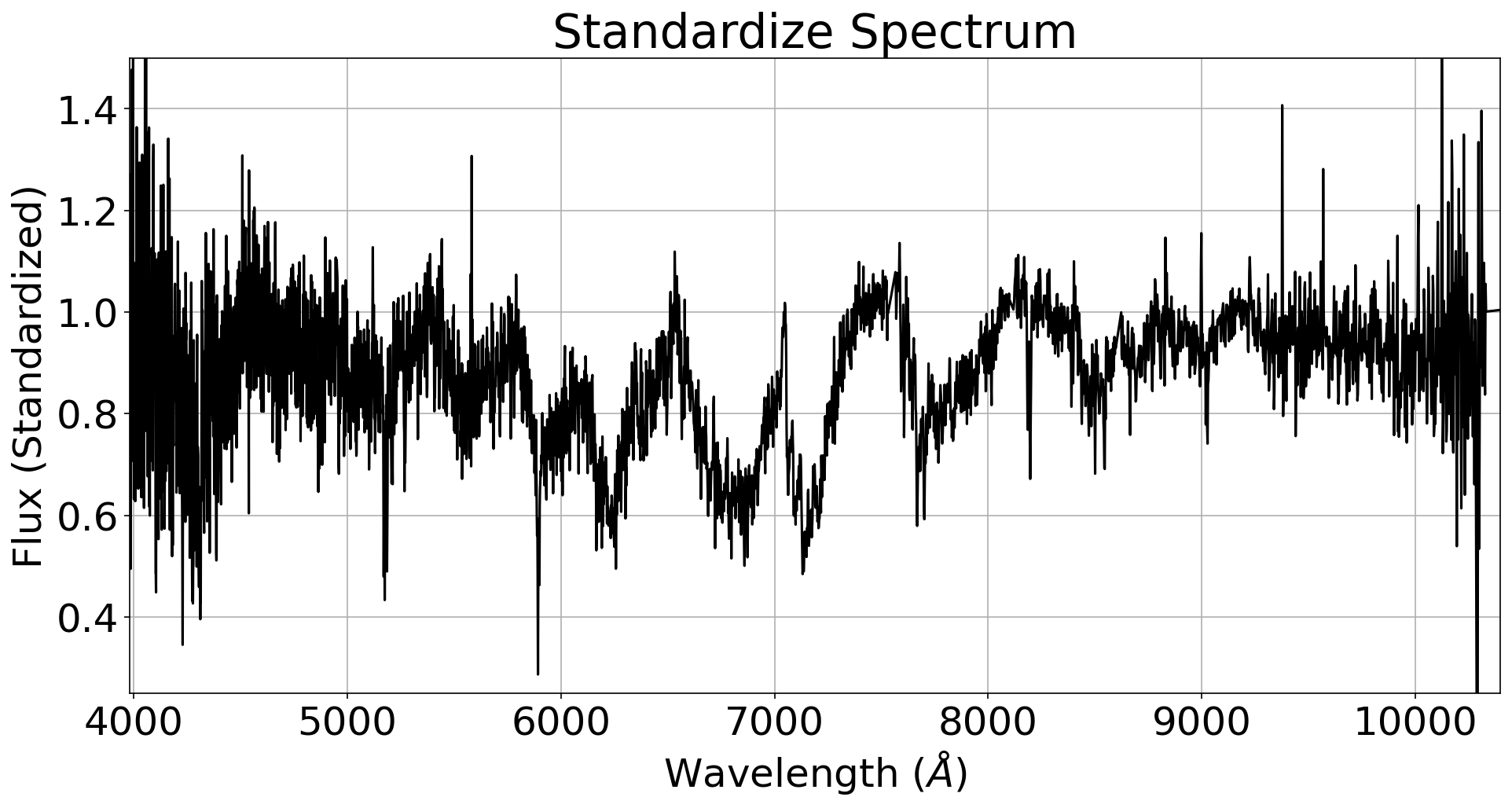}
	\caption{Plots demonstrating each step in our standardization method as described in Section \ref{sec:method_descript}.}
	\label{fig:method}
\end{figure}

As a result of the use of alpha shapes and the local polynomial regression, the Alpha Hulling Standardization method is highly flexible, but the set of hyperparameters that describe the above must be tuned to ensure the desired outcome. These will be described in \S\ref{sec:constrain_hyper} when we outline our optimization. 
In summary, the steps of the Alpha Hulling Standardization method are displayed in Figure \ref{fig:method} and are described as follows:
\begin{enumerate}
    \item \textbf{Reduce the Spectrum:} The method begins with a spectrum reduced with the BOSS pipeline \citep{Bolton2012, Dawson2013,morrison2025}. For all M dwarf spectra, we only consider the flux for $\lambda > 10^{3.6} \approx 3981 \ \text{\AA}$.
    \item \textbf{Apply Masking and Median Filter:} \added{Here we mask} the bad pixels and run a median filter of the spectrum. For the mask, we do a local sigma clipping where we run in a window size of $200\times 10^{-4} \ log(\AA)$ and mask any flux values that are greater than $5\sigma$ from the mean. We then apply a median filter to the non-masked pixels with a window size of $13\times 10^{-4} \ log(\AA)$.
    \item \textbf{Set Aspect Ratio:} We normalize the spectrum to match a set aspect ratio. This means that, the wavelength and flux data are normalized such that the ratio of the difference between the maximum and minimum flux value of the masked, median filtered spectrum, and the difference between the maximum and minimum wavelength value equals the specified aspect ratio. 
    \item \textbf{Find Alpha Shape:} Find the alpha shape of the median filtered spectrum normalized to our aspect ratio. The ``ball" of the alpha shape (with radius $1/\alpha$) is shown in blue \added{in Figure \ref{fig:method}}.
    \item \textbf{Identify Maximum Points of Alpha Shape:} Identify all the vertices along the surface of our alpha shape, as they define the top of our spectrum. Ideally, these will skip over the molecular bands. This will occur for alpha balls that are larger than the width of the molecular bands. If the ball is too small, they will fall into the bands and pick up points within them.
    \item \textbf{Perform Local Polynomial Regression:} Perform a local polynomial regression on the vertices along the surface of the alpha shape. Here we use a Gaussian kernel with some standard deviation. We would like to note here that we based the local polynomial regression code written for our method on \texttt{localreg} \citep{sigvald_marholm_2022_6344451} and made changes to decrease the runtime of the code.
    \item \textbf{Standardize Spectrum:} Divide the spectrum by the newly defined pseudo-continuum. In practice, the local polynomial regression calculates the pseudo-continuum in normalized flux and wavelength units (as defined by the aspect ratio). So, the pseudo-continuum must be re-scaled to absolute units first. After this, the standardized spectrum is produced by dividing the reduced spectrum by the re-scaled pseudo-continuum.
\end{enumerate}

\subsubsection{Constraints on Hyperparameters }\label{sec:constrain_hyper}

The Alpha \added{Hulling} Standardization method defined in Section \ref{sec:method_descript} requires a set of hyperparameters as input, which drastically change the shape of the final pseudo-continuum. These include:
\begin{itemize}
    \item \textbf{Aspect Ratio:} The aspect ratio of flux to wavelength used \added{to re-scale the} spectra.
    \item $\boldsymbol{\alpha}$: The inverse of the radius of the alpha \added{ball} used for the hulling of the spectra.
    \item \textbf{Kernel:} The kernel function used for the local polynomial regression.
    \item \textbf{Radius:} The size applied to the kernel function along the normalized (by the aspect ratio) wavelength axis.
    \item \textbf{Order:} The order of the polynomial for the local polynomial regression.
\end{itemize}
To find the optimal hyperparameters, we must have some evaluation of how well we are standardizing the spectra. Typically, one would utilize a subset of SDSS-V stars with known stellar parameters. If these parameters are well constrained and a good set of theoretical models exist, we could test our standardization method on both the observed and model spectra and evaluate how well they compare to one another.
As stated in the introduction, though, this work is currently incomplete for M dwarfs, so finding a set of labels that covers the full range of parameter space is currently impossible.

We choose a different approach. Instead, we start with a set of atmospheric models of M dwarfs to generate model spectra at the resolution of BOSS. Although we do not expect these to perfectly reproduce the features in the real M dwarf spectra, they still have the same overall shape and sets of features we expect for M dwarfs, namely, the locations and approximate widths of the oxides (TiO, VO) and hydrides (CaH, MgH), and as a result will serve as a good testing ground for our method. For this work, we use the BT-NextGen models \citep{Allard2011} accessed via the SVO Theoretical Spectra repository\footnote{\url{http://svo2.cab.inta-csic.es/theory/newov2/}} with the AGSS2009 \citep{AGSS2009} Solar chemical abundance. These models were computed using the NextGen code with updated line opacities \citep{BT-NextGen-1,BT-NextGen-2,BT-NextGen-3,BT-NextGen-4}. The ``BT" in the title refers to \cite{BT}, which provides a high-accuracy list of water lines. BT-NextGen does not incorporate the cloud models of BT-Settl \citep{BT-Settl} but covers a wider range in metallicity and approximately reproduces the broad molecular features seen in M dwarf spectra.
These models will then be used to create ``BOSS-like" spectra, such that after standardizing both the models and the BOSS-like spectra, we can see how well we can recover the original spectra in all regions of $T_{eff}$, $[Fe/H]$ and $[\alpha/Fe]$.

All the BT-NextGen theoretical spectra are convolved to a resolution of $R=2000$ and then sampled down to the wavelength grid of BOSS. We must now add a few things to this spectrum to mimic what we may observe with BOSS. Since M dwarfs in SDSS-V can be observed out to distances of $\sim 1$ kpc, reddening must be added to the model spectrum. We use the reddening law from \citet{Cardelli1989} and randomly assign a magnitude of visual extinction based on the distribution of M dwarfs in SDSS-V. Here, we select the M dwarfs that were targeted in DR18 and find the extinction of each source based on the dust map of \citet{green2018}. This distribution of visual extinctions was best fitted by a log-normal distribution with $ln(\mu) = 0.28$ and $\sigma = 1.54$. Finally, we add noise to the spectrum by assuming it is Gaussian and choosing a SNR from a uniform distribution with a minimum of 5 and a maximum of 60. When creating BOSS-like spectra for our optimization, we do not enforce a correlation between SNR and extinction. We do this as we want to prioritize probing the full range of possible spectra rather than having a strong correlation between the two parameters.

Reddening and Gaussian noise alone do not reproduce the flux calibration issues seen in Figure \ref{fig:ex_var_mdwarf}. To recreate these issues would involve modeling the response function of BOSS, adding in poorly subtracted sky lines, simulating incorrect flux calibration, etc. This would be quite a complex procedure, but we have devised a data-driven approach that we believe will produce a similar result. We begin by selecting all BOSS spectra where $BP - RP > 1.84$, $M_G >8.16$ \citep{pecaut2013}, $\pi > 2$ mas, the SDSS-V program is MWM, SNR$ >2$ and the target has $>3$ epochs of observations with a radial velocity measurement. In this selection, we include spectra reduced with version \texttt{v6\_1\_3} of the pipeline \citep{morrison2025} and observed before MJD $= 60562$ (2024/09/09). Next, we find all objects which have spectra observed at multiple epochs and create a set of difference spectra, which are the difference between all combinations of the epoch spectra for the target after they have been radial velocity corrected and normalized with the mean flux value for $7495 < \lambda < 7505  \ \AA$. These difference spectra should then track the variance due to the flux calibration issues we are seeing in the observed BOSS spectra. Some estimate of these difference spectra can be used to make the models look more like these observed spectra. 

\begin{figure*}
	\centering
	\includegraphics[width=0.8\textwidth]{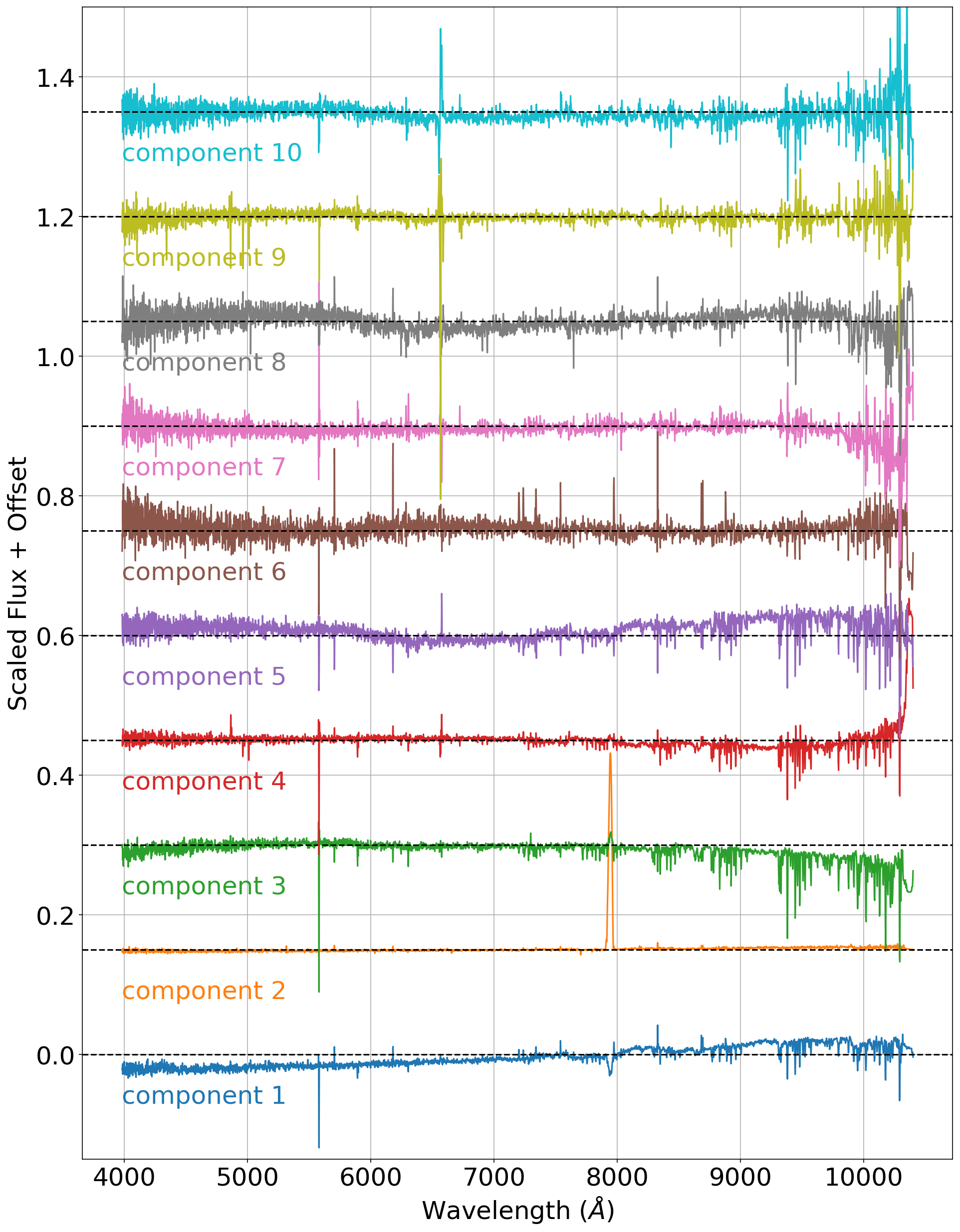}
	\caption{The first 10 components of the PCA performed on the difference spectra. Here we use reduced BOSS spectra from version \texttt{v6\_1\_3} of the pipeline where $BP - RP > 1.84$, $M_G >8.16$, $\pi > 2$ mas, the SDSS-V program is MWM, SNR$>2$ and the target has $>3$ epochs of observations with a radial velocity measurement. A difference spectrum is then the difference between all combinations of the epoch spectra for the target after they have been radial velocity corrected and normalized with a mean value for $7495 < \lambda < 7505  \ \AA$. The PCA is then performed with a total of 30 components.}
	\label{fig:pca}
\end{figure*}

To reproduce the BOSS-like variations due to poor flux calibration, we use a PCA analysis with 30 components to obtain the eigenspectra of the difference spectra, where the first 10 components are shown in Figure \ref{fig:pca}. These eigenspectra, when added to a model, would change the shape of the SED, add in random high-flux lines, and cause fringing effects at the edge of the chip. When we combine a random weighting of these PCA components with the reddening and noise above, we should then be able to reproduce a convincing BOSS-like spectrum from a model.

\begin{figure*}
	\centering
	\includegraphics[width=0.9\textwidth]{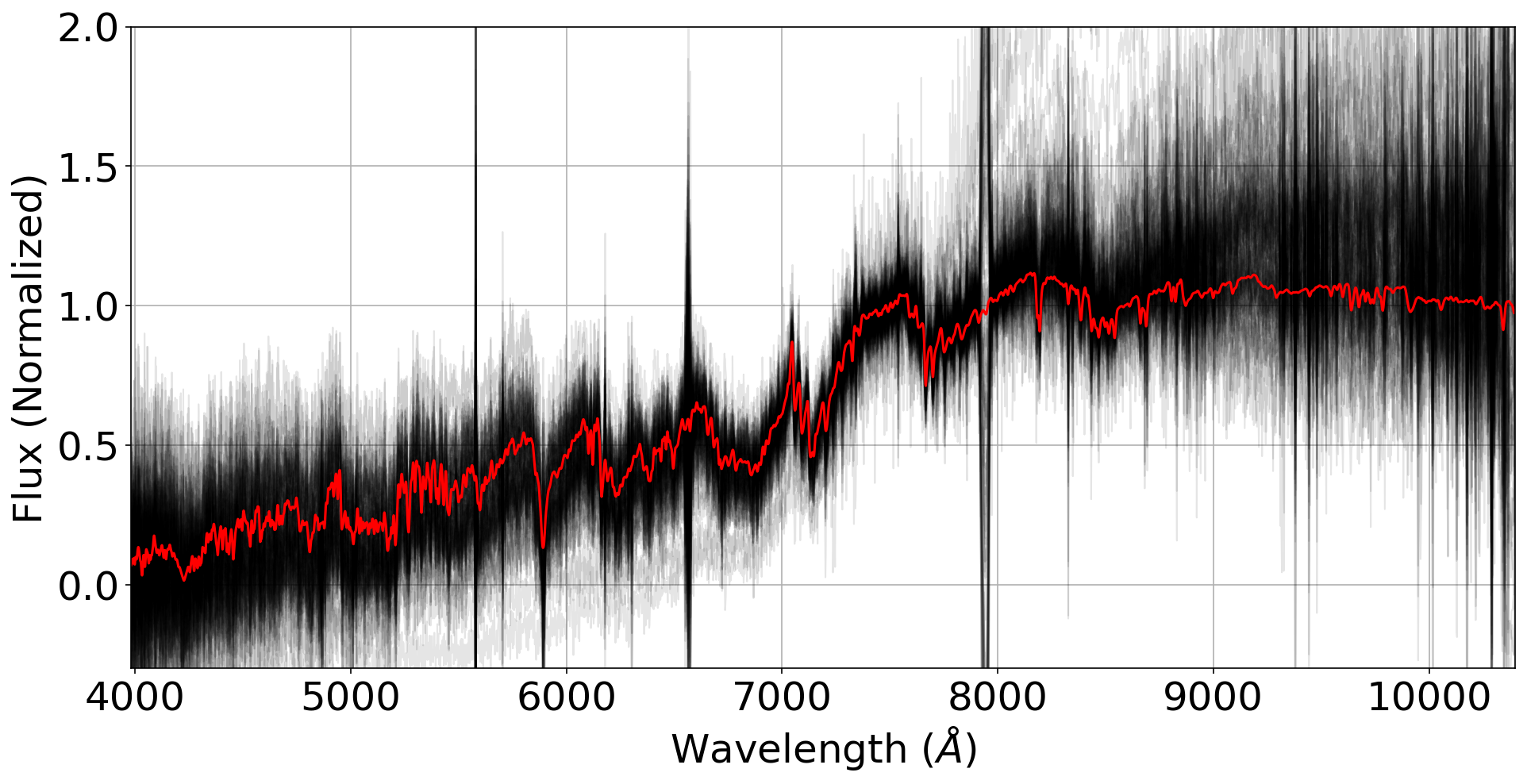}
	\caption{An example of a set of ``BOSS-like" spectra (black lines) generated by adding a random realization of a difference spectrum from our PCA bases, along with some random amount of extinction and noise to a smoothed BT-NextGen model (red line). In this figure, the black lines are 100 random realizations of such a BOSS-like spectrum for a BT-NextGen model with $T_{eff} = 3400$ K, $[Fe/H]=0$ dex and $[\alpha/Fe]=0$ dex. All spectra have been normalized based on the average flux value for $7495 < \lambda < 7505  \ \AA$.}
	\label{fig:sdss_like_ex}
\end{figure*}

Figure \ref{fig:sdss_like_ex} plots 100 realizations of a BOSS-like spectrum for a BT-NextGen model with $T_{eff} = 3400$ K, $[Fe/H]=0$ dex and $[\alpha/Fe]=0$ dex. The resulting BOSS-like spectra differ greatly from the original model, like the overall variance seen in the observed BOSS spectra in Figure \ref{fig:ex_var_mdwarf}. These BOSS-like spectra also have fringing issues at the red end of the spectrum, along with some large, poorly subtracted sky lines. We additionally see a few, highly reddened objects among the BOSS-like spectra, which greatly change the shape of the overall SED. Qualitatively, this process produces spectra that replicate those observed with BOSS and will serve as a perfect test for our method.

The final thing to consider is a model of the errors in our BOSS-like spectra, which will be important to get a realistic calculation of the $\chi^2$ between the models and the BOSS-like spectra for subsequent analysis. We use a Random Forest Regressor to predict the inverse variance (IVAR) of a spectrum based on its flux measurements. For a training set, we use the same spectra used to create the difference spectra above. We uniformly sample stars between $8< M_G < 20$ mag in bins of 2 mag. When there are enough stars available, we select 300 stars, but at the low-luminosity end of the sample we use as many stars as possible due to low numbers. We then train a regressor with 100 trees in the forest and find that the resulting regressor reproduces the IVAR well, where we see low IVAR near the edge of the chip, that sky lines are well masked and that the IVAR strongly weights strong molecular absorption features. This gives us the final piece we need to construct our BOSS-like spectra.

\subsubsection{Optimization of Hyperparameters }\label{sec:opt}

To find the ideal hyperparameters, we create 20 BOSS-like spectra for each BT-NextGen model, as described in Section \ref{sec:constrain_hyper}. Our BT-NextGen models span the range of $2300 \leq T_{eff} \leq 3900$ K in bins of 100 K, $-4 \leq [Fe/H] \leq 0.5$ dex in bins of $0.5$ dex and $0 \leq [\alpha / Fe] \leq 0.4$ dex in bins of $0.2$ dex. All models are for $log(g) = 5$. For models with $T_{eff} < 2600$ K, we only have models at Solar metallically. From the hyperparameters listed at the beginning of Section \ref{sec:constrain_hyper}, we choose to fix the kernel to a Gaussian function and the order of polynomial used in the local polynomial regression to a 3rd order. This was based on an initial, sparse grid search of hyperparameter space, where we found these choices have a plurality in their preference. So, fixing these helps reduce the dimensionality of the subsequent search.

For the remainder of the parameters ($\alpha$, radius and aspect ratio), we perform a grid search to find the ideal values. We impose the following boundaries for these parameters. For $\alpha$, we set the minimum to $1/5$ and the maximum to $1/0.02$. For radius, we set the minimum to the size of a pixel in our normalized space (i.e.~$1/4170$) and the maximum to the full extent of our normalized wavelength space (i.e.~$1$). Finally, for the aspect ratio, we set the minimum to 1 and the maximum to 10. An aspect ratio below 1 is possible (where the wavelength axis is spread further than the flux) but is not well behaved. The increasingly large gaps between flux measurements lead to an ill-defined alpha-shape, causing our method to fail. These bounds are summarized in Table \ref{tab:hyperpar_table}.

\begin{figure*}
	\centering
	\includegraphics[width=0.49\textwidth]{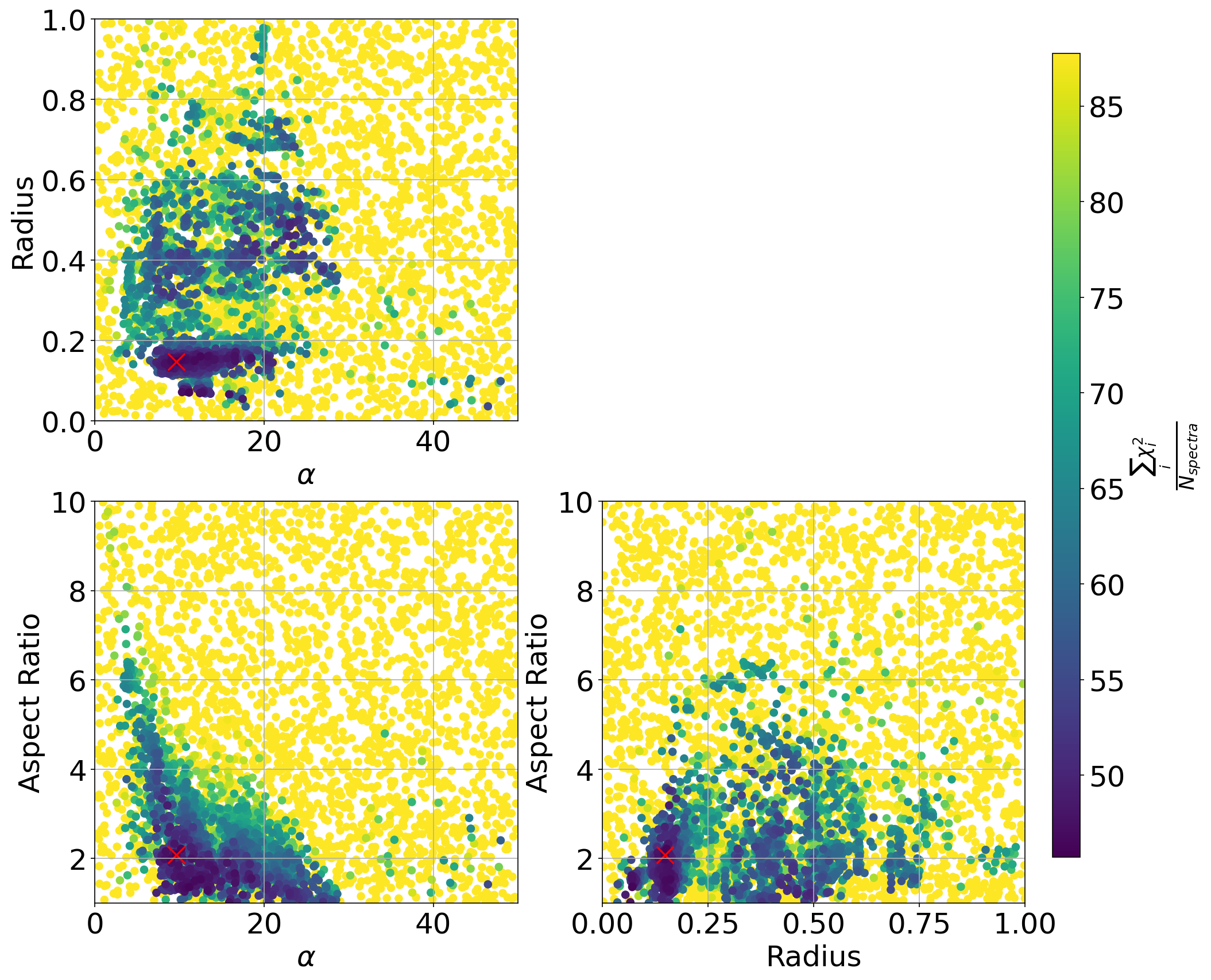}
    \includegraphics[width=0.49\textwidth]{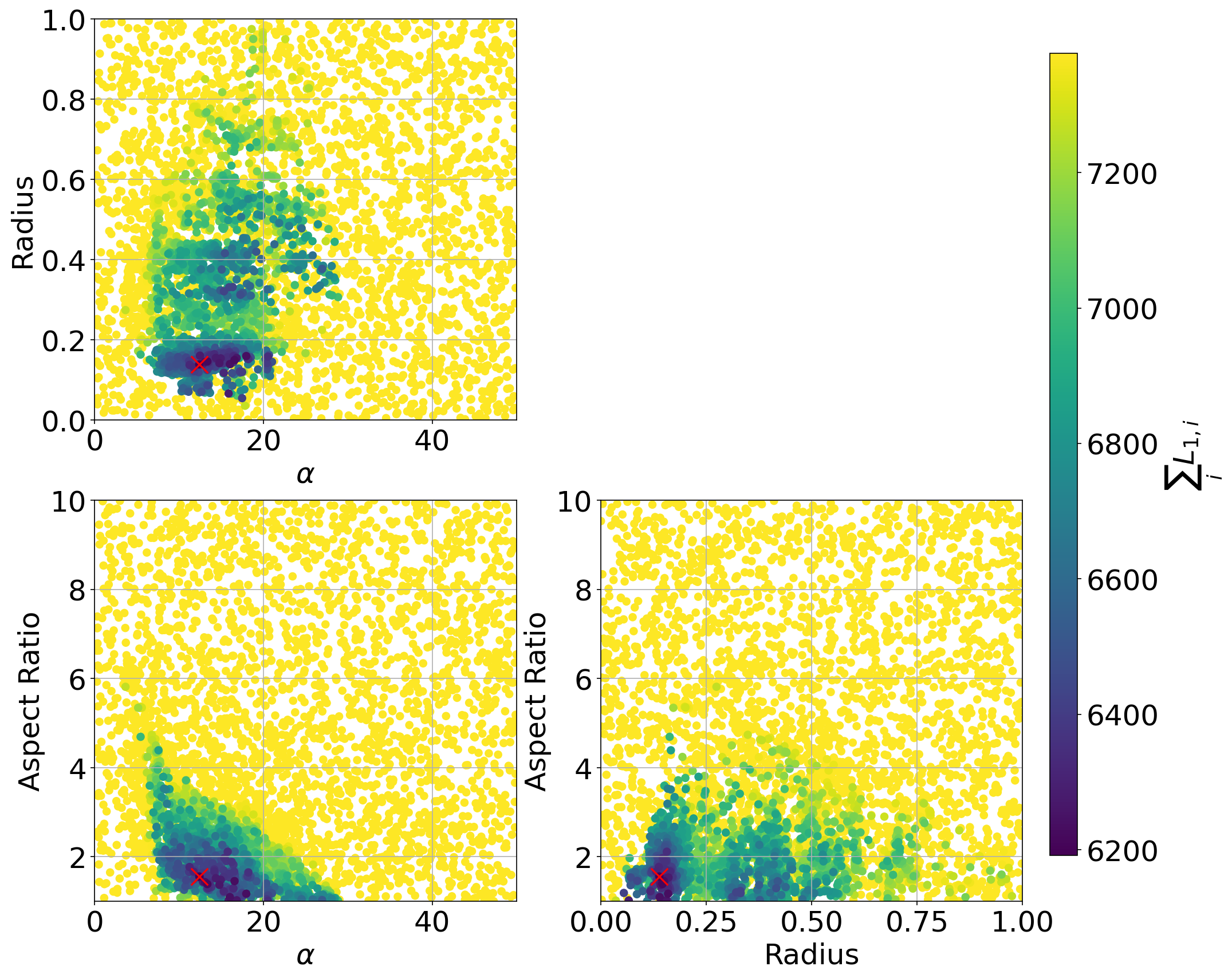}
	\caption{Results of the grid search for the hyperparameters $\alpha$ (inverse of size of alpha hull), radius (size of kernel used for smoothing in local polynomial regression) and aspect ratio (spectrum is normalized with aspect ratio of flux to wavelength). The grid search was first run over the full bounds of the parameters (Table \ref{tab:hyperpar_table}) in a course selection of 2,500 randomly selected points. With these hyperparameters, we then standardized all BT-NextGen models and BOSS-like spectra, and calculated the $\chi^2$ and $L_1$ norm for each spectrum normalized with those parameters. After this, we selected the course grid points that were either in the minimum 5\% of $\sum_i \chi_i^2$ or $\sum_i L_{1,i}$ values, and resampled between all of these adjacent grid points for another 2,500 parameters, and evaluated the $\chi^2$ and $L_1$ for each BOSS-like spectrum at these points. The corner plots show the resulting $\sum_i \chi_i^2 / N_{spectra}$ (left) and $\sum_i L_{1,i}$ (right) evaluations, where it should be noted the colorbar cuts off at the 40\% of the values. The red cross shows the global minimum in each case.}
	\label{fig:grid_results}
\end{figure*}

We uniformly sample 2,500 random points from this grid for the initial search. With these hyperparameters, we then standardize all BT-NextGen models and BOSS-like spectra, and calculate the $\chi^2$ and $L_1$ norm for each spectrum standardized with those parameters. Here, $\chi^2$ is the variance-weighted squared difference between a model and a BOSS-like spectrum, and $L_1$ is the distance (in units of bins the models are sampled on) between the true parameters of the model and the parameters of the model matched by the minimum $\chi^2$ between the models and the BOSS-like spectrum. This grid search was coarse, so we elect to resample the space with a finer grid. To do this, we select the coarse grid points that were either in the minimum 5\% of $\sum_i \chi_i^2$ or $\sum_i L_{1,i}$ values (i.e.~the sum of the $\chi^2$ and $L_1$ norm for all spectra standardized with those hyperparameters), and resample between all of these adjacent grid points for another 2,500 parameters. It should be noted that in this resampling procedure, we removed the highest 5\% of $\chi^2$ values from the sum due to outlier BOSS-like spectra with consistently extreme $\chi^2$ values. We finally evaluate the $\chi^2$ and $L_1$ values of all spectra at these fine grid points. Figure \ref{fig:grid_results} shows the result of this grid search. The two corner plots show $\sum_i \chi_i^2 / N_{spectra}$ (left) and $\sum_i L_{1,i}$ (right) for all models and BOSS-like spectra standardized using these hyperparameters, where we again removed the highest 5\% of $\chi^2$ values from the sum due to outlier BOSS-like spectra with consistently extreme $\chi^2$ values.
This was occurring because of very large flux values in these spectra, leading to $\chi^2$ many orders of magnitude above the mean. 

The grid search demonstrates that the radius is very well constrained, especially in the $\sum_i L_{1,i}$ surface, and matches the width of the larger molecular bands in normalized flux units. This makes sense, as we would expect a well-working standardization method to smooth the pseudo-continuum over the molecular bands to reduce any higher-order features in this region. Generally, a wide range of aspect ratios seem to work well with our method, but an aspect ratio near 1.5 seems to work best. Finally, we see that in the $\sum_i \chi_i^2 / N_{spectra}$ surface, there are multiple $\alpha$ values that produce good, standardized results. We see that there is an additional minimum for large values of $\alpha$ (small alpha \added{ball}). This causes some of the vertices along the top of the hull to be within the molecular bands, which leads to a pseudo-continuum that is like a polynomial fit through the entire spectrum. We see that this secondary minimum does not appear in the $\sum_i L_{1,i}$ surface, though, meaning that while it reduces the $\chi^2$ between the models and spectra, it is more difficult to recover the stellar parameters. This difference is key and will be a metric of how well different standardization methods perform in \S\ref{sec:compare_alt_methods}.

Overall, the global minima of the $\sum_i \chi_i^2 / N_{spectra}$ and $\sum_i L_{1,i}$ surfaces are similar, but not exactly the same. The biggest differences are in $\alpha$ and the aspect ratio, which differ by $\sim5\%$. Because of this, we will balance the two with a regularization term such that we look for the minimum in $\sum_i \chi_i^2 / N_{spectra} + \lambda \sum_i L_{1,i}$. We chose $\lambda = 0.01$, as this gives about equal weight to $\chi^2$ and $L_1$ norms. Our final hyperparameters are $\alpha = 12.48$, radius $= 0.16$ and aspect ratio $= 1.53$, as summarized in Table \ref{tab:hyperpar_table}.
Along with the fixed Gaussian kernel and 3rd order polynomial for the local polynomial regression, these are the parameters we will use for the subsequent analysis of our method. 

The Alpha Hulling Standardization method above is available for use via the code in our GitHub repository\footnote{\url{https://github.com/imedan/mdwarf_contin/tree/1.0.0}}. By default, the code uses the optimized hyperparameters found here. An example of how to use this code with a BOSS spectrum of an M dwarf is shown in Appendix \ref{app:example}.

\begin{table*}[!t]
  \centering
  \caption{The hyperparameters \added{of the various standardization methods are shown}. The \textit{Grid Search Prior} column shows the span of values searched in the optimization, and the \textit{Optimum} column shows the \added{best value(s) for the} method.} 
  \begin{tabular}{|l |c | c |c|}
\hline
       \textbf{Method} & \textbf{Hyperparameter} & \textbf{Grid Search Prior} & \textbf{Optimum} \\
        \hline
        \hline
         Alpha Hulling & Aspect Ratio & $\mathcal{U}(1,10)$ & \added{1.53} \\
         \cline{2-4} & $\alpha$ & $\mathcal{U}(1/5, 50)$ & \added{12.48} \\
         \cline{2-4} & Kernel & Gaussian & Gaussian \\
         \cline{2-4} & Radius & $\mathcal{U}(1/4170, 1)$ & \added{0.16} \\
         \cline{2-4} & Order & 3 & 3 \\
      \hline
      \hline
      Polynomial & Order & $\mathcal{U}(1, 20)$ & 7 \\
      \hline
      \hline
      GISIC & $\sigma$ & $\mathcal{U}$(1 \AA, 200\AA) & $22.51$\ \AA \\
      \hline
      \end{tabular}
  \label{tab:hyperpar_table}
\end{table*}

\subsection{Alternative Methods}\label{sec:alt_methods}

We have described in detail our own methodology for standardizing M dwarf spectra, but it is by no means the only way to approach this problem. There are several available methods that one could use. Here, we describe three of these alternative methods that we will compare to ours. This is not an exhaustive list, but instead a short, representative sample that will cover a range of methods, from simpler to complex (like ours).

The simplest method possible is constant standardization. Here, we assume that the flux calibration of the BOSS spectra is precise, so we can simply find a region of the spectrum that is close to continuum and use that average flux to standardize the spectra. In fact, in the \texttt{astra} data analysis pipeline for DR19 (version \texttt{0.6.0}), \textsc{MDwarfType}, which does the template matching for BOSS M dwarfs, this method is utilized. As can be seen from Figure \ref{fig:ex_var_mdwarf} though, this does not produce a consistent result. So, we expect that this approach will not be able to standardize the BOSS-like spectra in the subsequent analysis.

Another approach is to simply increase the order in which we are fitting the pseudo-continuum. Here we can fit the spectra with a higher-order polynomial to account for the changes in the shape of the SED and standardize the spectra. In \citet{LASPM_pipeline}, their approach was to use a fifth-order polynomial, which minimizes the differences between the models and the spectra. 
There is the question of \textit{which} polynomial order would best standardize our spectra. To evaluate this, we calculate the $\chi^2$ and $L_1$ norm for the BOSS-like spectra, as discussed in the previous section, for polynomial orders 1 through 20. When fitting the polynomial, we apply the same masking, median filtering, and aspect ratio as we do with our method using the optimal parameters found in \S\ref{sec:opt}. We find a global minimum in the $\sum_i \chi_i^2 / N_{spectra}$ at an order of 12, and an order of 7 for $\sum_i L_{1,i}$. There are local minima at order 7 for the $\sum_i \chi_i^2 / N_{spectra}$ and order 11 for $\sum_i L_{1,i}$ as well. Because we want to prioritize the recovery of the correct parameters after standardization, we fit the pseudo-continuum with this alternative method using a 7th order polynomial in the subsequent analysis.

The final method we consider is similar in nature to our method, but with a different approach. Gaussian Inflection Spline Interpolation Continuum (GISIC\footnote{\url{https://pypi.org/project/GISIC/}}) standardizes the spectra by performing Gaussian smoothing of the spectra and identifying molecular bands based on the numerical gradient of the flux. The ``continuum" points outside the molecular bands are then interpolated with a cubic spline to get the pseudo-continuum. So, like our method, GISIC attempts to identify the top of the spectrum outside the molecular bands and fit these points to standardize the spectra. GISIC has been used to standardize spectra like those considered here. \citet{yoon2020} used GISIC to normalize both MARCS models and spectra of carbon-enhanced metal-poor (CEMP) stars, as GISIC helped with the standardization around the strong molecular carbon bands in their medium resolution optical spectra. 
GISIC has also been used to standardize M dwarf spectra by the CARMENES \citep{carmenes1, carmenes2} team when training deep learning models to predict M dwarf parameters \citep{Passegger2020, mas2024}. There is one free parameter to vary in GISIC, the width of the Gaussian used for the filtering. Like our previous methods, we complete a grid search for $1 < \sigma < 200 \ \AA$. Before applying GISIC, we run the same masking as we do for our method. From this search, we find a global minimum in $\sum_i \chi_i^2 / N_{spectra}$ of $\sigma = 1 \ \AA$ and $\sigma = 22.51 \ \AA$ for $\sum_i L_{1,i}$. For the $\sum_i \chi_i^2 / N_{spectra}$ surface, a small $\sigma$ is optimal as it completely removes all features from the spectrum, meaning it can always well match the model spectrum. This results in a very high $\sum_i L_{1,i}$ though, as all characteristic features of the spectra are removed, making it difficult to recover the true parameters. We note that in the $\sum_i \chi_i^2 / N_{spectra}$ surface though, the minimum is very broad and extends to $\approx 60 \ \AA$. Like the polynomial, we choose to use the minimum of $\sigma = 22.51 \ \AA$ for the subsequent analysis, as it will better recover the correct parameters after standardization.

\section{Results}\label{sec:results_overall}

\subsection{Performance with Simulated Data }\label{sec:results}

In the previous sections, we described both our Alpha Hulling method and a few alternative methods for standardizing M dwarf spectra. The ideal set of hyperparameters for these methods was found based on how well we could match the BOSS-like spectra to the original model spectra they were created from (Table \ref{tab:hyperpar_table}). In this section, we will compare each of these methods with their ideal set of hyperparameters.

\begin{figure*}
	\centering
	\includegraphics[width=\textwidth]{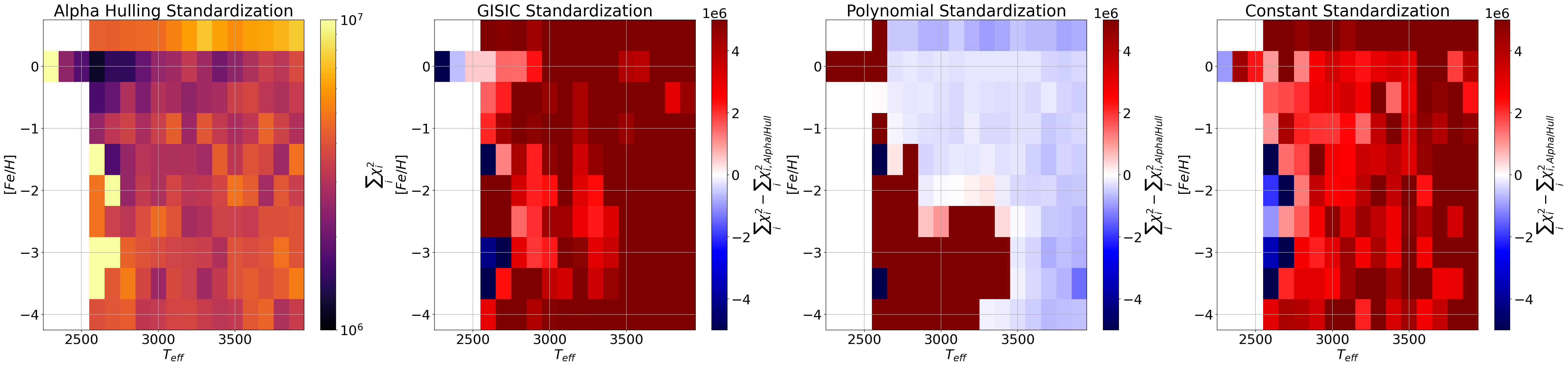}
	\caption{The sum of the $\chi^2$ values $(\sum_i \chi_i^2)$ between the all BOSS-like spectra and BT-NextGen models for each grid point in $T_{eff}$ and $[Fe/H]$ when spectra are standardized with our Alpha Hulling method (left panel). For this sum, we only consider the \added{lower} 90th percentile of $\chi^2$ values within a grid point. The other three panels show the difference between this sum for the other standardization methods and our method. The colormap is set such that red regions are where the standardization performs better with our method, and blue is when it performs better for the other method. Our method generally performs better for large regions of parameter space. Both GISIC and the constant standardization performs better for the lowest temperature, metal-poor stars, and the polynomial standardization is slightly preferred for high temperature stars and \added{some} lower temperature, metal-rich stars.}
	\label{fig:results_compare}
\end{figure*}

The initial comparison we make is to see how well the original, model spectrum compares to its manipulated, BOSS-like spectrum. To do this, we compute the sum of the $\chi^2$ values $(\sum_i \chi_i^2)$ between the standardized, BOSS-like spectra and their true model in the bins of $T_{eff}$ and $[Fe/H]$ shown in Figure \ref{fig:results_compare}. This statistic should be sensitive to whether features of the original standardized spectra are still present in the corresponding BOSS-like spectra. In this sense, a lower $\sum_i \chi_i^2$ value means that the standardization method produces BOSS-like spectra that more closely resemble their original models.
The far-left panel shows the $\sum_i \chi_i^2$ for our method, where we see comparable $\sum_i \chi_i^2$ values across parameter space. We see that the $\sum_i \chi_i^2$ values are larger on average for some metal-poor cool M dwarfs. We also see that the $\sum_i \chi_i^2$ is on average smaller for the cooler solar metallicity spectra.

In the other panels of Figure \ref{fig:results_compare}, we show the difference in the $\sum_i \chi_i^2$ between the other standardization methods and our method. Here, the color scaling means that the Alpha Hulling Standardization $\sum_i \chi_i^2$ is smaller when red and larger when blue. For GISIC, we see that our method produces a better fit between the BOSS-like spectra and the models for all regions of parameter space. The exception is for cool, metal-poor M dwarfs. The polynomial method produces lower $\sum_i \chi_i^2$ values for all regions, except for the metal-poor cool objects, where it performs significantly worse than our method. Finally, the constant standardization method performs better and worse in regions like GISIC.

With only this statistic it would seem that polynomial standardization is the most precise method. But this is not the full picture. In practice, a standardization method could overfit the spectrum and ``standardize out" essential information that would be needed to, say, differentiate between metal-poor and metal-rich M dwarfs. Indeed, a standardization method can drastically reduce the difference between the original and BOSS-like spectra if both are transformed into completely straight lines. So, rather than finding whether each method recovers the original spectrum, it is necessary to characterize how accurately each method recovers a spectrum's original labels. For the remainder of this section, we will focus on how well stellar parameters are recovered in a grid search.

\subsubsection{Alpha Hulling Standardization Method}

\begin{figure*}
	\centering
	\includegraphics[width=\textwidth]{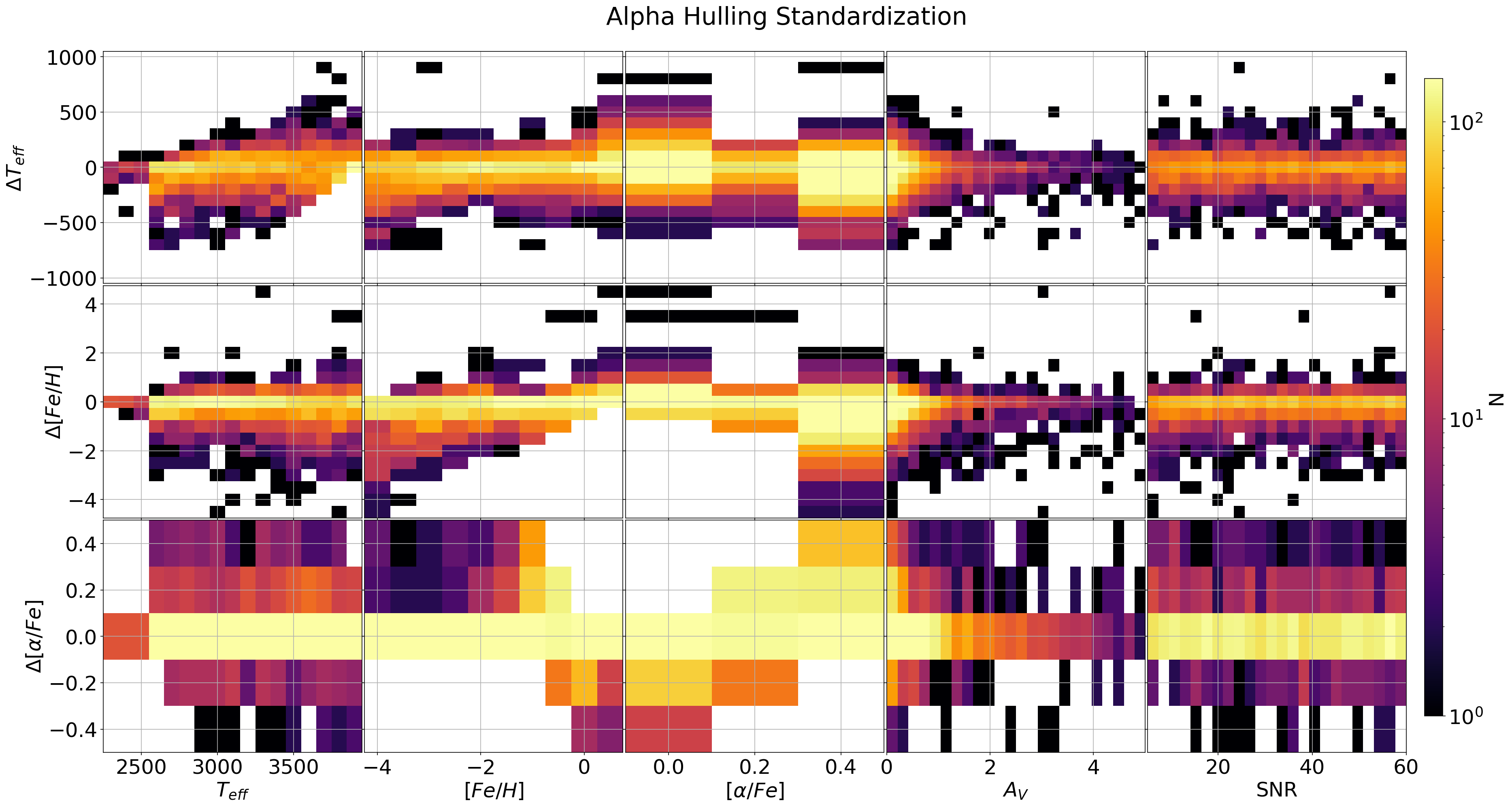}
	\caption{The best stellar parameters are found by minimizing the $\chi^2$ between the standardized BOSS-like spectra to the standardized BT-NextGen models. This imitates a typical template matching that would be performed on SDSS-V spectra. The above shows the difference between the true and best fit parameters vs.~the true parameters of the spectrum used to generate the BOSS-like spectra.}
	\label{fig:results_params}
\end{figure*}

For Alpha Hulling Standardization, we looked at the difference in the true stellar parameters of the spectrum and what is recovered from a $\chi^2$ minimization between the BOSS-like spectra and the models. Figure \ref{fig:results_params} shows the difference between the true and best fit parameters vs.~the true parameters of the model used to generate the BOSS-like spectra. We generally see that the best fits match the correct parameters of the model with a tight distribution. Specifically, for $T_{eff}$, $[Fe/H]$ and $[\alpha/Fe]$ we find an overall $1\sigma$ scatter of the difference to be $159.1$ K, $0.632$ dex and $0.092$ dex, respectively. For most parameters in most regions of parameter space, the differences are normally distributed around zero. There are some exceptions, such as for the metal-poor objects where it seems that the recovered metallically can be overestimated from the template matching. We see a similar effect at the low temperature end of our models as well where we see a slight skew towards $\Delta T_{eff} < 0$ for spectra of $T_{eff} \sim 2700$ K. It is especially reassuring to see that the differences in parameters do not vary as a function of the added visual extinction and signal-to-noise ratio. Our method can recover the true parameters regardless of either of these factors. 

We can improve the grid search for some parameters by only evaluating the $\chi^2$ for $\lambda > 6500 \ \AA$. This is done by calculating the pseudo-continuum on the full, valid region of the spectrum ($\lambda > 3981 \ \AA$), but only using $\lambda > 6500 \ \AA$ to calculate $\chi^2$ in the grid search. This results in an overall $1\sigma$ scatter of the difference of $132.8$ K, $0.834$ dex and $0.099$ dex for $T_{eff}$, $[Fe/H]$ and $[\alpha/Fe]$, respectively. The average $[Fe/H]$ is larger because it creates a skew for hotter M dwarfs. For $T_{eff} < 3500 \ K$, the scatter drops to 0.5 dex for $[Fe/H]$. Although this improves the results for the grid search, we do not think this cut is advisable. For our validation tests, which we will discuss in Section \ref{sec:valid}, the blue end of the spectrum is needed for better separating stars on the HR diagram. This is supported by the larger bias in the metallicity determination for hotter stars when removing the blue part of the spectrum.

\subsubsection{Comparing to Alternative Methods}
\label{sec:compare_alt_methods}

\begin{figure*}
	\centering
    \subfloat[Alpha Hulling Standardization]{\includegraphics[width=0.85\textwidth]{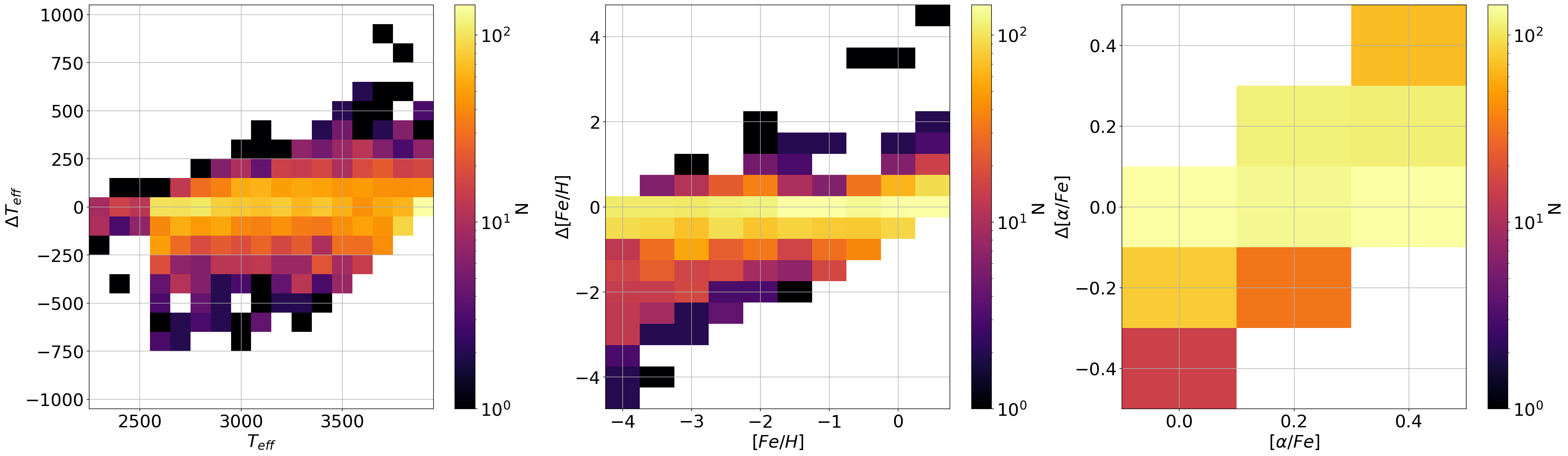}}\\
	\subfloat[GISIC Standardization]{\includegraphics[width=0.85\textwidth]{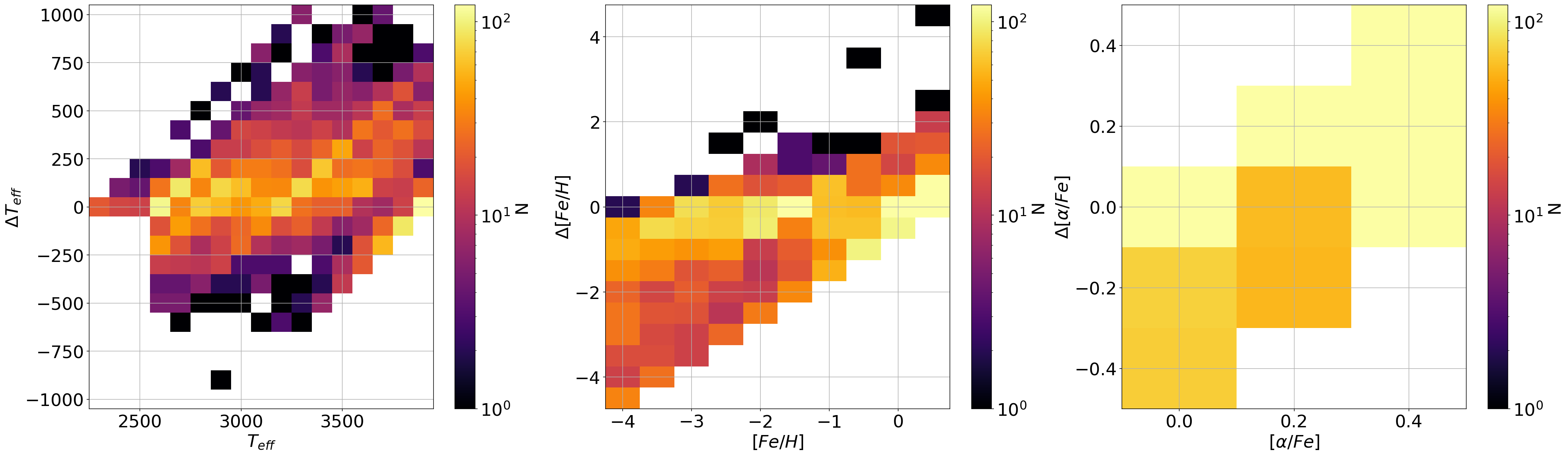}}\\
    \subfloat[Polynomial Standardization]{\includegraphics[width=0.85\textwidth]{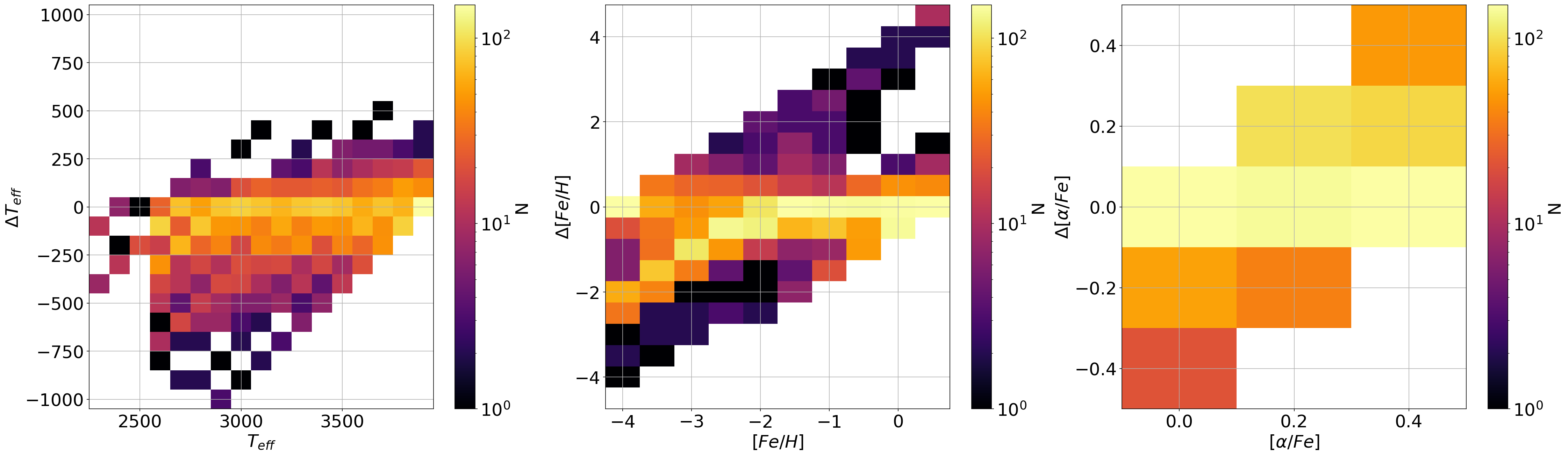}}\\
    \subfloat[Constant Standardization]{\includegraphics[width=0.85\textwidth]{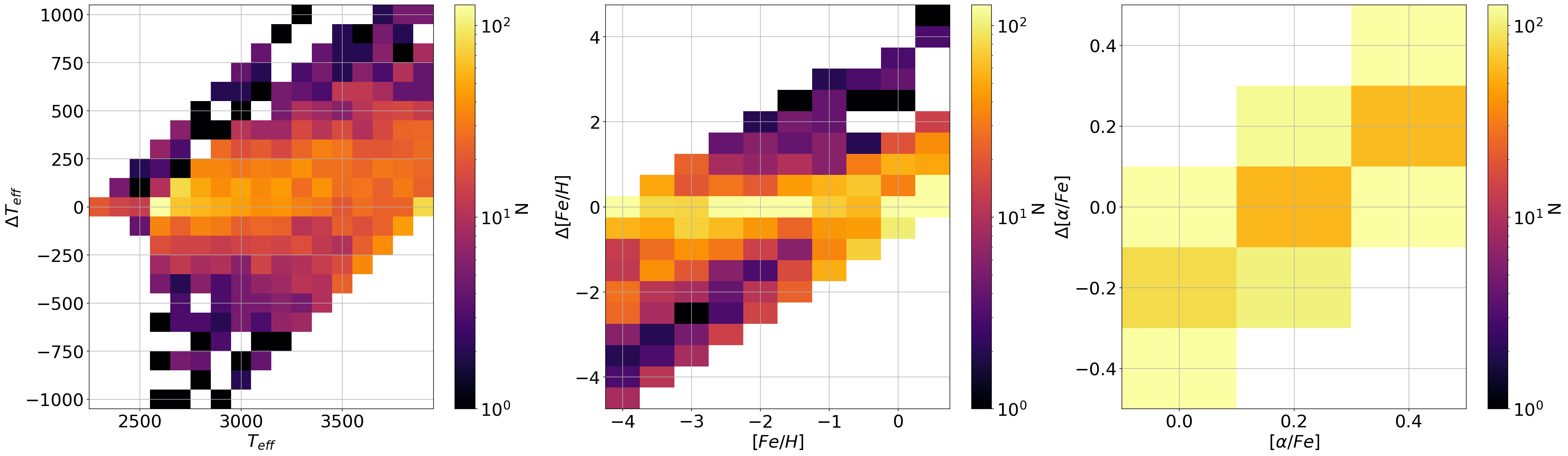}}\\
	\caption{The best stellar parameters are found by minimizing the $\chi^2$ between the standardized BOSS-like spectra to the standardized BT-NextGen models. This process imitates a typical template matching that would be performed on SDSS-V spectra. The above shows the difference between the true and best fit parameters vs.~the true parameters of the spectrum used to generate the BOSS-like spectra \added{for the various methods.} Overall, the scatter is low when the spectra are standardized with our method, and we see very few covariances in the parameters.}
	\label{fig:results_params_compare}
\end{figure*}

Next, we compare the recovery of the true parameters for the three alternative methods. Figure \ref{fig:results_params_compare} shows the difference between the true and best fit parameters vs.~the true parameters of the spectrum used to generate the BOSS-like spectra for the various methods. The method with the \added{worst} performance is constant standardization (Figure \ref{fig:results_params_compare}, bottom row). The variance is especially large in $T_{eff}$ ($\sigma \approx 298$ K) due to a combination of the extinction and the overall flux calibration issues greatly changing the shape of the SED. In a constant standardization scheme, the shape of the SED will largely drive the $T_{eff}$ fit, causing the very large errors seen here.

Polynomial standardization (Figure \ref{fig:results_params_compare}, third row) on the other hand appears to perform significantly better on average. However, the polynomial method struggles at lower temperature models, where we see that $T_{eff}$ is typically overestimated from the $\chi^2$ minimization. Similarly, we see that the metal-poor stars have overestimated metallicities to a greater degree than what is found with our Alpha Hulling method.

Finally, GISIC standardization (Figure \ref{fig:results_params_compare}, second row) performs poorly in regions different from the polynomial method. Primarily, GISIC seems to be underestimating $T_{eff}$ for the higher-temperature M dwarfs. Metal-poor stars also have overestimated metallicities to a larger degree than both the Alpha Hulling Standardization and polynomial methods. Finally, the GISIC results seem to have many grid edge effects where parameters at the edge of the parameter space are more likely to be recovered.

\begin{figure*}
	\centering
    \includegraphics[width=\textwidth]{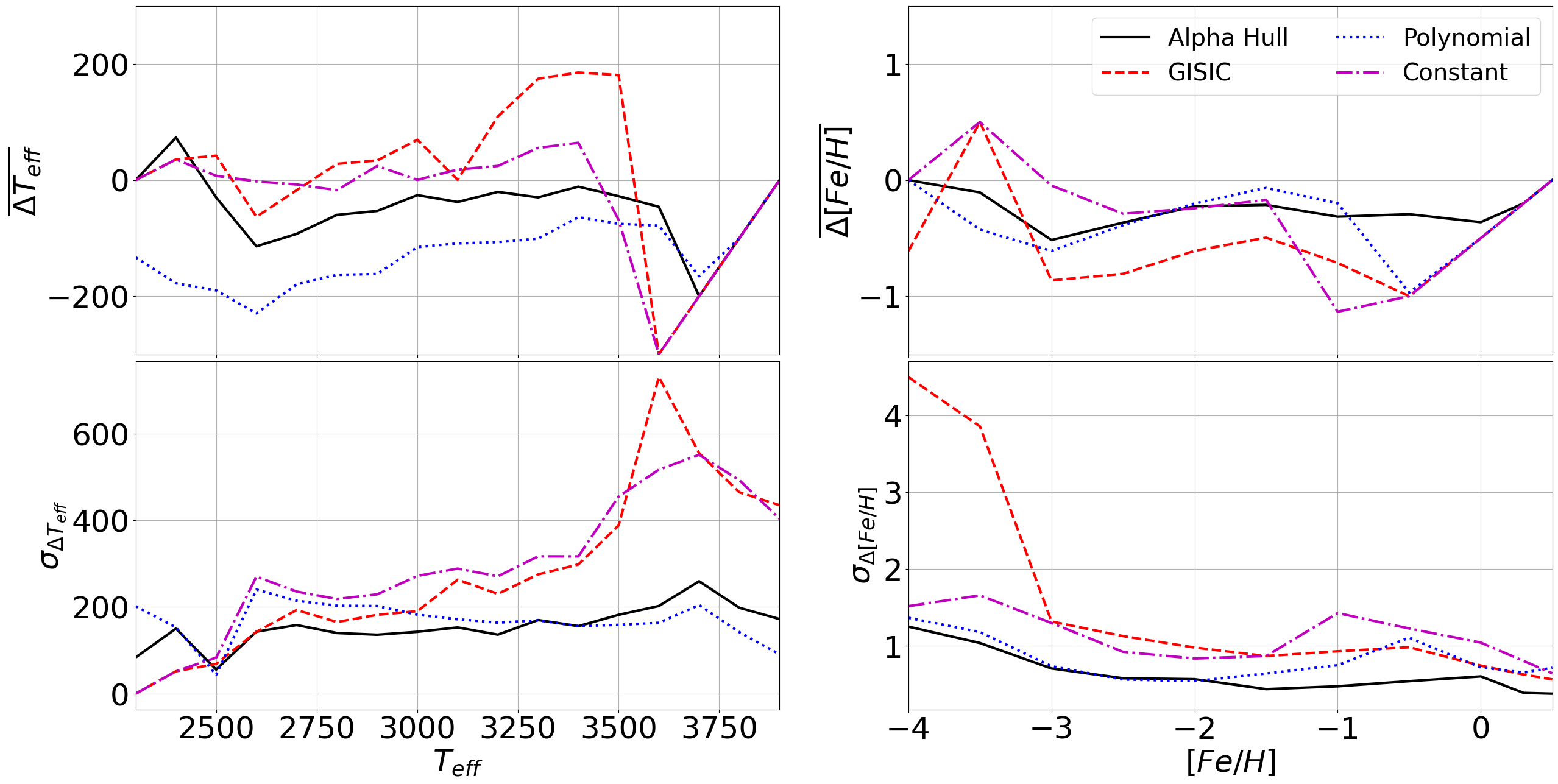}
	\caption{The mean (top row) and standard deviation (bottom row) \added{from a fit of a truncated normal distribution to} the difference between the true and best fit parameters from our grid search when BOSS-like spectra and BT-NextGen models are standardized using the various methods. This demonstrates the level of bias (top row) and scatter (bottom row) for each method, where we find that our Alpha Hulling method generally outperforms or produces similar results as the alternative methods.}
	\label{fig:results_params_bias}
\end{figure*}

To better quantify the above, we wanted to calculate the accuracy and precision of the results from the grid search when using the different standardization methods. This can be probed by calculating $\overline{\Delta T_{eff}}$ and $\overline{\Delta [Fe/H]}$, which measures the level of accuracy in the results, and $\sigma_{\Delta T_{eff}}$ and $\sigma_{\Delta [Fe/H]}$, which measures the level of precision. In both cases, values closer to zero indicate that the stellar parameters are being recovered more accurately and precisely, respectively, for a given method. One issue is that we are performing a grid search with edges to find the best parameters. So, when the true parameters are at the edge of the grid, the resulting $\Delta$ distribution cannot be normal. In fact, it is a truncated normal distribution that is cut off at the maximum distances set by the grid edges. So, as a function of $T_{eff}$ and $[Fe/H]$, we fit a truncated normal distribution to calculate $\overline{\Delta T_{eff}}$ and $\overline{\Delta [Fe/H]}$, and $\sigma_{\Delta T_{eff}}$ and $\sigma_{\Delta [Fe/H]}$ for each method (Figure \ref{fig:results_params_bias}).

In terms of the $T_{eff}$ recovered, we find that our Alpha Hulling method outperforms the alternative methods. In term of accuracy ($\overline{\Delta T_{eff}}$), we find that our method greatly outperforms the polynomial standardization, especially at lower temperatures. Additionally, our method outperforms the GISIC and constant standardization at higher temperatures. Although our method is comparable in accuracy in regions outside this, in terms of precision ($\sigma_{\Delta T_{eff}}$), our method offers lower scatter. This is especially true at higher temperatures, where the GISIC and constant methods perform much worse. Compared to polynomial standardization, the results are like our Alpha Hulling method in terms of precision, but, again, recall that our method greatly outperforms in terms of accuracy compared to the polynomial method. In general, these results demonstrate that our Alpha Hulling Standardization is always on par or outperforms all alternative methods in recovering $T_{eff}$.


There is a similar picture for $[Fe/H]$. In Figure \ref{fig:results_params_bias}, for $-3 < [Fe/H] < -1.5$ we find that our Alpha Hulling method performs similarly to the polynomial and constant methods in terms of accuracy ($\overline{\Delta [Fe/H]}$) but produces better results than the GISIC standardization. For the more metal-poor stars $([Fe/H] \leq -3)$ and more metal-rich stars $([Fe/H] > -1.5)$, we find that our method is closer to zero (better accuracy) than \added{all} the alternative methods. In terms of precision ($\sigma_{\Delta [Fe/H]}$), our method consistently outperforms both GISIC and constant standardizations, especially at lower metallicities. However, when compared to the polynomial standardization, the results are roughly equal for the precision achieved, except at the metal-rich end, where it has a slightly higher scatter. Again, we see that there was an improvement in $[Fe/H]$ accuracy using our method compared to the polynomial for $[Fe/H] \leq -3$ and $[Fe/H] > -1.5$, so our Alpha Hulling method still offers some advantage. Overall, these results demonstrate that our Alpha Hulling method can recover stellar parameters in a grid search, which offer either comparable or improved accuracy and precision compared to alternative methods for most regions of parameter space.


\subsubsection{Summary}

Although some of the alternative methods have a lower overall $\chi^2$ for some regions of parameter space (see Figure \ref{fig:results_compare}), this is not necessarily a beneficial result. Indeed, one can imagine if a method removed all features from a spectrum such that the standardized spectrum was only a line with Gaussian noise, the $\chi^2$ would indeed be low, but all information about the stellar parameters would be lost. The main goal of our method is to reduce this $\chi^2$, while maximizing the information retained within the standardized spectra. 
The fact that the standardization method used influences the recovery of the stellar labels, as shown in  Figure \ref{fig:results_params_bias}, leads to two key conclusions. First, the standardization method used can have a substantial effect on \textit{both} the precision and accuracy of recovered stellar parameters. Second, by using a method that is well tuned for the M dwarfs, like our Alpha Hulling method, both the bias and scatter can be reduced. This demonstrates the ability of our Alpha Hulling method and how it better standardizes spectra of M dwarfs for analyses like template matching or generative model training.

\subsection{Validation: Performance with SDSS BOSS Data}\label{sec:valid}

\subsubsection{Alpha Hulling Standardization Method}

\begin{figure*}
	\centering
	\includegraphics[width=\textwidth]{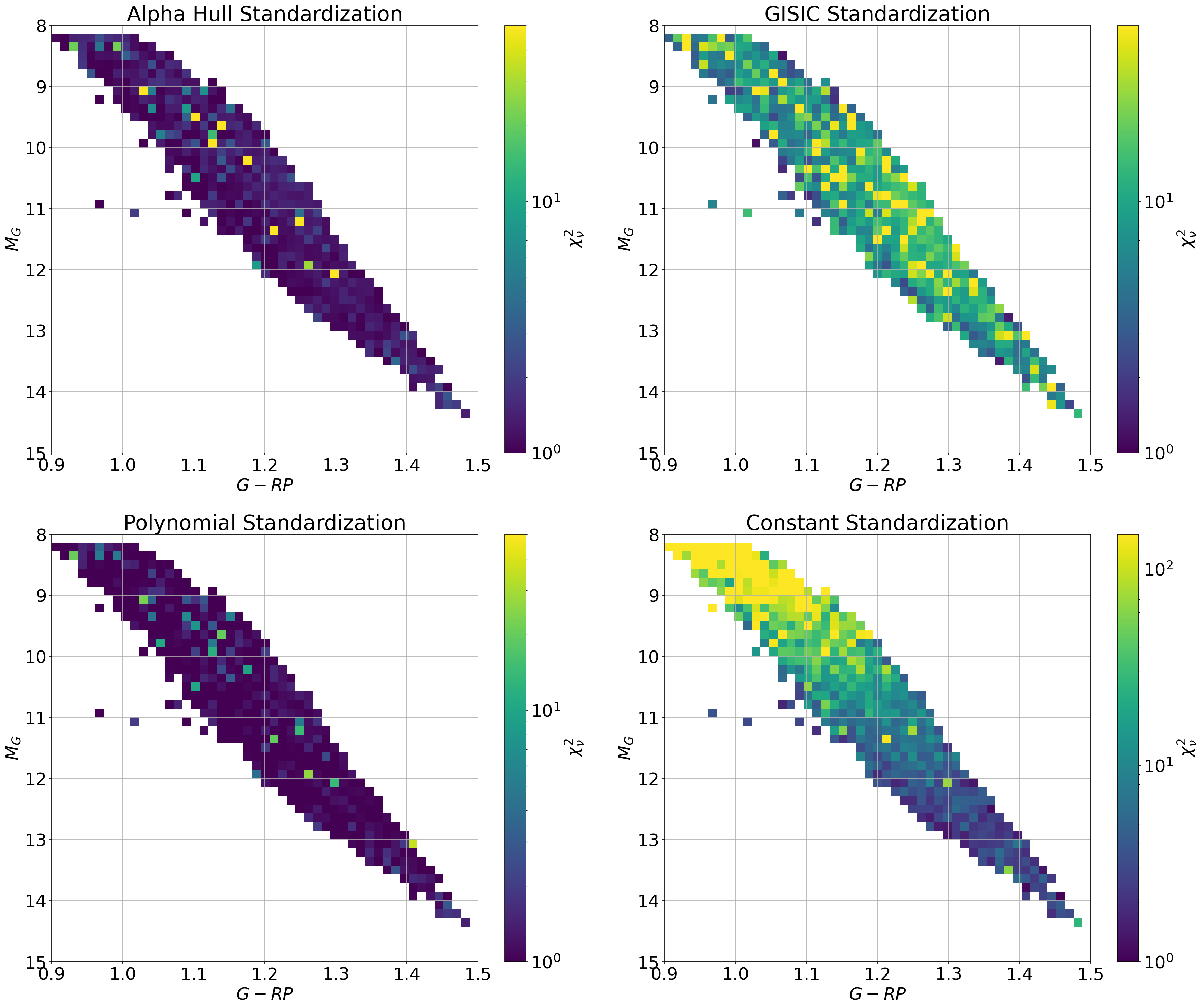}
	\caption{HR diagram of BOSS spectra from SDSS-V, where each grid point shows the $\chi^2_\nu$ between the median spectrum in that bin and all spectra standardized with a given method. In the above, only BOSS spectra within 100 pc, with a SNR$> 5$ and determined to be true single stars by \cite{way2025} were included in the above. Additionally, only grid points with $>5$ spectra are shown. All spectra were reduced using \texttt{v6\_2\_0} of the BOSS pipeline and include data up until MJD $= 60715$ (2025/02/09).}
	\label{fig:valid_compare}
\end{figure*}

\begin{figure*}
	\centering
	\includegraphics[width=\textwidth]{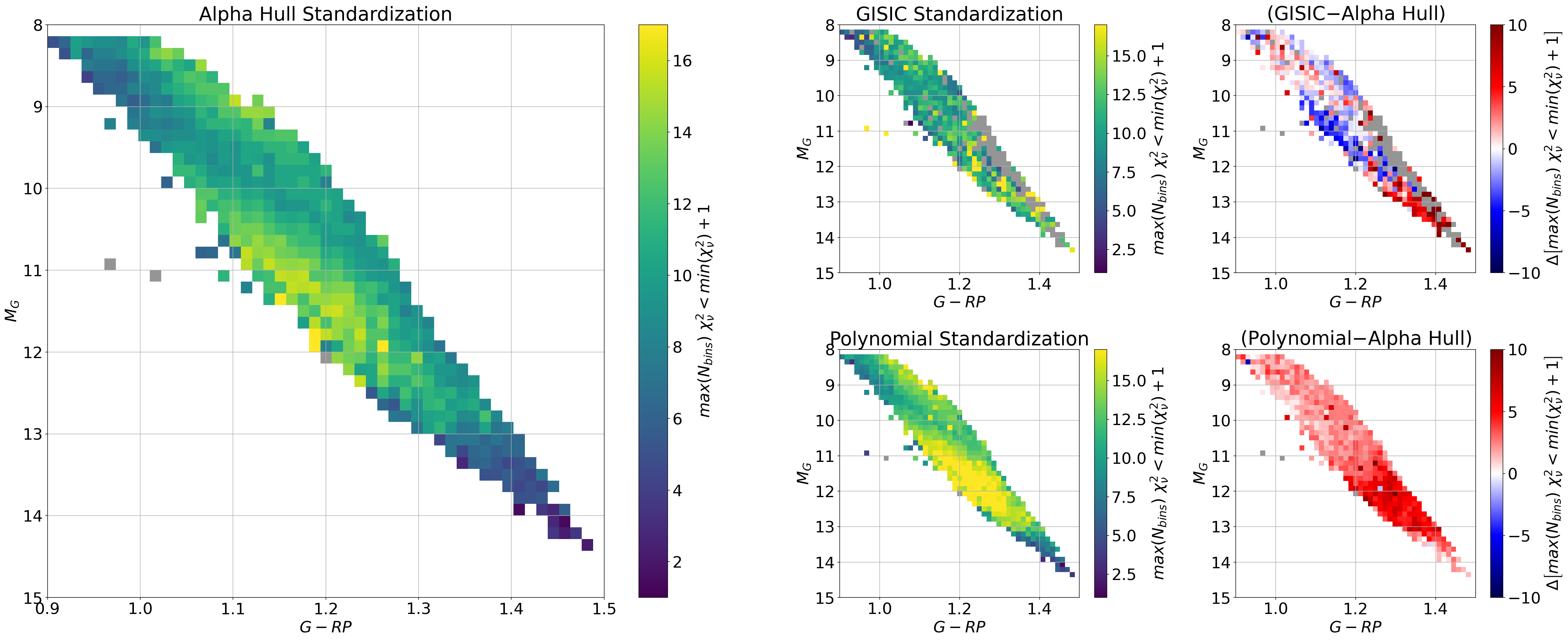}
	\caption{HR diagram of BOSS spectra from SDSS-V, where each grid point shows the maximum distance (in number of grid points) where $\chi^2_\nu < min(\chi^2_\nu) + 1$. Here, for each grid point the $\chi^2_\nu$ is calculated between all standardized spectra in that grid point and then for each median, standardized spectra in all other grid points. This allows for the visualization of the $1\sigma$ level variance in regions across the HR diagram. For bins where the distance between the HR diagram bin for the spectra and the bin of the $min(\chi^2_\nu)$ is greater than the maximum distance where $\chi^2_\nu < min(\chi^2_\nu) + 1$, the bin is grayed out as the solution is not statistically significant. The different panels in the above show this variance level for the different standardization methods. The far-right column of plots then shows the difference in the variance level between the other methods and our method. For these plots, the colorbar indicates that the variance is smaller with our method when red. In all the above plots, only BOSS spectra within 100 pc, with a SNR$> 5$ and determined to be true single stars by \citet{way2025} were included in the above. Additionally, only grid points with $>5$ spectra are shown. All spectra were reduced using \texttt{v6\_2\_0} of the BOSS pipeline and include data up until MJD $= 60715$.}
	\label{fig:valid_compare_varriance}
\end{figure*}

Up to this point, we have examined the results of our Alpha Hulling method in terms of faux, BOSS-like data that we have generated from model spectra. To validate our method, we really want an understanding of how it performs with actual BOSS spectra. In this section, we outline a data-driven validation of our method to demonstrate how the above results are reflected in what we will see with the real BOSS data.

For this validation, we will use the BOSS M dwarf spectra within 100 pc, with a SNR$> 5$ and with Gaia DR3 data \citep{gaiadr3}. There is a degeneracy between the spread in metallicity and the unresolved, overluminous binaries in this region of the \textit{Gaia} CMD. We select the single stars in this sample using the methods from \cite{way2025}. This work identifies unresolved, equal-mass binaries using their \textit{Gaia} XP spectra. It first predicts the absolute magnitude of a source and identifies binary stars as too bright for a given XP spectrum. This works because an equal-mass binary will emit twice the flux but should emit the same SED as their single-star counterparts.
With this set of criteria, all stars for a small region on the HR diagram should then have very similar $T_{eff}$ and $[Fe/H]$, as they are not polluted by binaries and the stars are too close to be affected by reddening. So, once the spectra are standardized, all the spectra within a bin should be very similar to each other.

To test this, Figure \ref{fig:valid_compare} shows a HR diagram of these BOSS spectra where each grid point shows:
\begin{equation}
    \chi^2_\nu = \sum_i\sum_{\lambda}{\frac{(f_{i,\lambda} - \overline{f_{\lambda}})^2}{\sigma_{i, \lambda}^2}} \times \nu^{-1}
\end{equation}
where $f_{i,\lambda}$ is the standardized flux of the ith spectrum in the pixel on the HR diagram at some wavelength $\lambda$, $\overline{f_{\lambda}}$ is the median flux at some wavelength $\lambda$, $\sigma_{i, \lambda}^2$ is the variance in the flux of the ith spectrum in the pixel on the HR diagram at some wavelength $\lambda$, and $\nu$ is the degrees of freedom defined as the sum of the number of spectra in the bin times the length of the wavelength grid for each spectrum. We qualitatively see a similar result as in Figure \ref{fig:results_compare}. Overall, with our Alpha Hulling method we find that the $\chi^2_\nu$ is low across the entire HR diagram. We see a few pixels with particularly high $\chi^2_\nu$, which are typically due to a small number of spectra in that bin with unusually large flux values, greatly increasing the overall $\chi^2_\nu$. Besides this, though, most pixels have $\chi^2_\nu$ around 1, indicating that all spectra are like each other across the HR diagram.

\subsubsection{Comparing to Alternative Methods}

We also calculated the $\chi^2_\nu$ for spectra standardized with alternative methods, as shown in Figure \ref{fig:valid_compare}. GISIC performs worse compared to our Alpha Hulling method by an order of magnitude and has a larger $\chi^2_\nu$ for most regions. The constant standardization is up to two orders of magnitude worse and is displayed on a separate color bar to show its dynamic range. In particular, we see a trend in which the highest-mass M dwarfs have the highest $\chi^2_\nu$ values. This is because the higher mass stars are intrinsically brighter and more likely to have higher SNR spectra. With a higher SNR, the changes in the shape of the SED are more apparent and will be consistently much greater than the noise. As constant standardization does not address changes in the SED shape, this is why there is such a strong correlation with luminosity. Finally, we see that polynomial standardization performs slightly better than our method and has a $\chi^2_\nu$ around 1 for most of the HR diagram. This result is expected based on the comparison in Figure \ref{fig:results_compare}. We do not see the regions where we expect the polynomial method to perform worse (metal-poor, cool M dwarfs), as these types of M dwarfs are not numerous enough within 100 pc to appear in this plot.

While the polynomial method produces standardized spectra that look like the median spectrum in a bin, this does not imply that the standardization method retains information about the star. As stated in the previous section, the more crucial factor is how well you can recover a spectrum with the same parameters among all the other spectra with differing parameters. This is something that we can also test with our SDSS validation set. To do this, we now calculate the $\chi^2_\nu$ surface across the HR diagram. What this means is that, for a particular bin, you calculate the $\chi^2_\nu$ between those spectra and all median spectra for every bin in the HR diagram. Then, you can find the $1\sigma$ variance region by considering the maximum distance (in number of grid points) where $\chi^2_\nu < min(\chi^2_\nu) + 1$. This is essentially the region of the HR diagram where, statistically, the spectra all look similar to a $1\sigma$ level of confidence. This does work under the assumption that the $\chi^2_\nu$ \added{surface} is normally distributed and the covariance in two-dimensions is spherical on the HR diagram, so one value can well describe the variance. This is likely not the case, as some $\chi^2_\nu$ surfaces likely have a covariance where each dimension has its own general variance value. Additionally, this statistic only works well when the minimum $\chi^2_\nu$ is in the same bin as the spectra considered. If the minimum is farther away than the $1 \sigma$ region, then the result is not statistically significant. With these assumptions in mind, we want the standardized spectra to look distinct from others in different regions of the HR diagram, so we can have a more robust template match or generative model resulting from said spectra. So, we want this metric to be low, such that standardized spectra within a bin only look similar to a small region of the HR diagram surrounding them.

Figure \ref{fig:valid_compare_varriance} shows the result of this analysis. As a note, for all panels in Figure \ref{fig:valid_compare_varriance}, bins where the distance between the HR diagram bin for the spectra and the bin of the $min(\chi^2_\nu)$ is greater than the maximum distance where $\chi^2_\nu < min(\chi^2_\nu) + 1$ are grayed out as the solution is not statistically significant. The left panel of the figure is for our Alpha Hulling standardization, where we see that the maximum distance is relatively low across the HR diagram, with some regions being larger, particularity in the metal-poor region. If we compare this statistic to the GISIC and polynomial standardization, however, we see a very different result. For both the GISIC and polynomial methods, we see that the maximum distance is much larger for the relatively metal-poor, cool M dwarfs, such that these would be hard to fit when standardized with these methods. The far-right column of Figure \ref{fig:valid_compare_varriance} shows the difference in the statistic between these alternative methods and ours. These results echo the levels of accuracy and precision we saw from the BOSS-like spectra in Figure \ref{fig:results_params_bias}. In Figure \ref{fig:results_params_bias} we found that GISIC has much poorer precision for metal-poor stars and the hottest M dwarfs. In Figure \ref{fig:valid_compare_varriance} we find that for both regions of the HR diagram, it is more difficult to discern M dwarf spectra from their neighbors when the spectra are standardized with GISIC compared to our method. This would lead to higher levels of scatter in a grid search. For the higher-mass, metal-rich M dwarfs GISIC does perform better by this metric, though we do note this is a region of the HR diagram where there are relatively few M dwarfs. Additionally, for some of the stars with $M_G\sim11$, GISIC performs better. Though, for many bins in this region, the minimum in the $\chi^2_\nu$ is significantly far from the bin in question. This is due to the larger variance in the spectra within a bin, as shown by the elevated $\chi^2_\nu$ values in Figure \ref{fig:valid_compare}.

We see that the polynomial standardization performs worse in \added{nearly all} regions of the HR diagram, though the cause is different. From Figure \ref{fig:results_params_bias}, we found that the polynomial performed worse in terms of accuracy but was comparable in terms of precision. This result also manifests itself in Figure \ref{fig:valid_compare_varriance}, but in a different \added{form than} GISIC. For the polynomial, the reason why the maximum distance metric is so high for lower temperature regions of the HR diagram is that the $\chi^2_\nu$ surface is typically skewed to higher temperatures and breaks the assumption of being spherically symmetric in two dimensions. This means that there would be a bias to more likely confuse an M dwarf spectrum with locations on the HR diagram of higher temperature rather than lower temperature for stars with $M_G > 11$\added{. This} would manifest in a bias like what is seen in Figure \ref{fig:results_params_bias}.

\subsubsection{Summary}

Overall, these validation results demonstrate that our method can provide standardized spectra that produce similar looking spectra
for M dwarfs of similar stellar parameters and, importantly, can be well distinguished from other M dwarfs of differing parameters. These results reflect the trends we saw with our tests of the BOSS-like spectra, so we should expect similar improvements in terms of accuracy \textit{and} precision when our Alpha Hulling method is used. We want to emphasize that both improvements are crucial and will produce 
more robust results when our standardization method is used with future 
template matching or parameter estimation via generative models for 
these M dwarfs.


\section{Discussion}\label{sec:discuss}

\subsection{Limitations of Method}

There are a few limitations to the Alpha Hulling Standardization method that mean it cannot be applied universally to other medium-resolution optical spectra. Because the optimal hyperparameters depend on the geometry of the spectra in this specific range, they cannot be directly transferred to another survey. This is for a few reasons.

First, in the code for the method, the median filtering is based on a fixed $log(\AA)$ width, that is a multiple of the BOSS wavelength grid spacing. For differently spaced wavelength bins or for lower-resolution spectra, this would need to be adjusted, or spectra would need to be sampled to the wavelength grid of BOSS, if possible, before using the method as written for this paper.

Second, even if spectra from different surveys are on the same wavelength grid, if they cover a different wavelength range, the resulting pseudo-continuum will not be the same. When drawing the alpha shape, the vertices will always be placed at the first and last pixels of the spectrum. This could have a significant effect on the local polynomial regression. For example, if the spectrum started in the middle of an absorption band. The pseudo-continuum would be required to pass through the bottom of the band and would look much different from our current solution (which approximates a thermal-like continuum).

Overall, this means that the results and default parameters for the code that accompany this paper should only be used for the SDSS-V BOSS spectra. If this method were to be used for another survey, it would only be valid if 1) the spectra were sampled on the same wavelength grid as BOSS and 2) the flux covered the same wavelength range. If these two conditions are not satisfied, then the hyperparameters must be re-tuned to fit the survey. To avoid this, the spectra from the survey must be modified to match the BOSS spectra. This is possible, as we show below in a comparison of the BOSS and DESI M dwarf spectra.

\subsection{Future Directions}

In the future, this method could help facilitate the comparison of spectra between multiple spectroscopic surveys when training machine learning models. This should be possible if the spectra in both surveys 1) are of the same resolution and 2) cover the same wavelength range. There will likely not be complete overlap for both points, but there may be ways to mitigate this.

As an example, let us consider the optical spectra from DR1 of the 
Dark Energy Spectroscopic Instrument \citep[DESI;][]{DESI}. The DESI Milky Way program \citep{DESI_MW} targets stars within 100 pc like SDSS-V. DESI can observe stars at fainter magnitudes with higher signal-to-noise ratios, meaning that the combination of DESI and BOSS spectra would be advantageous for the study of local stars.

To understand how feasible this comparison would be, we examine the M dwarfs that have spectra in both DESI DR1 and SDSS DR19. The DESI spectra are higher resolution than the BOSS spectra and cover a different wavelength range. To place the spectra on the same wavelength grid, we linearly interpolate the higher-resolution DESI flux and errors to the wavelength grid of the BOSS spectra.  DESI spectra cover 3600--9800 \AA, while BOSS covers a wider 3600--10,400 \AA. To make a direct 1-to-1 comparison, the continuum must be calculated over the same wavelength range. So, we only calculate the pseudo-continuum for flux values $<9,800$ \AA \ for \textit{both} the BOSS and DESI spectra. With the way the code is written, the aspect ratio will remain the same despite this change. This is because, by default, the flux and wavelength data are normalized relative to the wavelength bounds used for the tuning of the hyperparameters (i.e.~$3981 < \lambda < 10,400$ \AA). So, this will allow us to find the pseudo-continuum without changing this hyperparameter.

\begin{figure*}
	\centering
	\includegraphics[width=\textwidth]{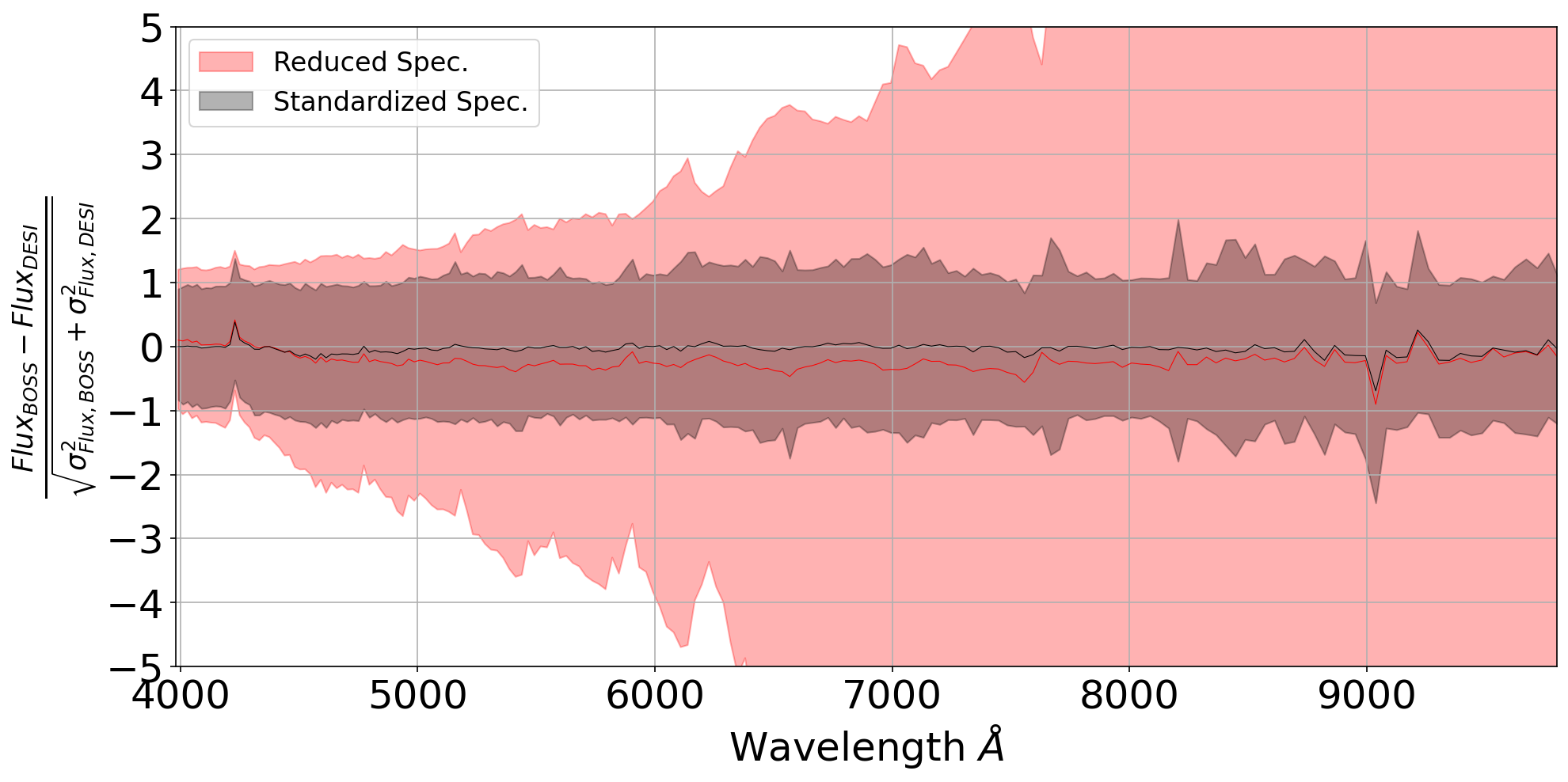}
	\caption{Comparison of spectra of 5,370 stars in both DESI DR1 and SDSS DR19. The plot shows the running median (solid lines) and 16th/84th percentile (shaded region) of the difference in the fluxes relative to the errors. The red is the difference in the reduced, flux-calibrated spectra directly from the pipelines of the respective surveys. The black is the difference in the standardized spectra using our Alpha Hulling method.}
	\label{fig:desi_compare}
\end{figure*}

Figure \ref{fig:desi_compare} shows the difference between the 5,370 stars in both DESI DR1 and SDSS DR19 relative to the flux errors. When using the flux-calibrated spectra directly from their respective pipelines (red), there are significant differences in the spectra of the same star. Such deviations are unlikely to be astrophysical and must be related to the issues in flux calibration. When using our Alpha Hulling method to standardize the spectra (black), we see that the DESI and SDSS spectra agree with each other, and the differences are within the errors of the flux measurements. This comparison demonstrates that a robust standardization method is key to combining spectra from complimentary surveys. Tools like our Alpha Hulling standardization method allow for this and will help facilitate detailed studies of the local population of M dwarfs.

Finally, we believe that this work will be important in facilitating machine learning applications for the optical M dwarf spectra, especially at low signal-to-noise ratio. In their discussion of attenuation bias, \citet{ML_bias} describes how noisier data causes machine learning models to underestimate extreme values. This poses a problem because the spectra of the faint M dwarfs are generally noisier than their high-mass counterparts, and their wide range of metallicity is interesting for several reasons. 
It is striking how similar the attenuation bias in Figure 1 of \cite{ML_bias} is to recent predictions of metallicity for low-mass stars in LAMOST \citep{zhang_m_dwarf_spec_TD, original_SLAM_Mdwarfs, LAMOST_empirical_Mdwarfs}, SDSS, and Gaia XP spectra \citep{qu_XP_phot}. 
However, as stated before, the spectra of M dwarfs are extremely reactive to differing abundance, even leading to large changes in photometric color. The abundance information is in the optical spectra, but we may be using the wrong tool in discriminative analysis. A generative approach is likely to yield much better constraints on stellar abundance, yet requires careful consideration about the standardization of the spectra. Our analysis in this paper offers a way to quantify how much inherent stellar information is kept when performing this standardization and paves the way for possible generative models.

\section{Summary and Conclusions}\label{sec:conclusions}

SDSS-V will produce an incredibly rich and diverse dataset due to hardware advances (i.e.~the FPS). Particularly in the optical, MWM will observe a large number of M dwarfs, accounting for nearly every 1-in-3 spectra observed with BOSS for MWM. After examining these spectra from the first half of the survey, we identified issues in the flux calibration from BOSS, meaning that some kind of standardization is needed to utilize this large amount of data in subsequent analyses.

In this work, we adapt an alpha hull standardization method inspired by an \'echelle spectral normalization technique \citep{alpha_shape_1, alpha_shape_2}, but tailor the hyperparameters of the method to work well with medium-resolution optical spectra of M dwarfs. To accomplish this, we adapt BT-NextGen models to look like SDSS spectra by adding reddening, noise and representative changes in the SED based on a set of difference spectra of M dwarfs. These resulting ``BOSS-like" spectra look incredibly similar to real BOSS spectra and serve as an ideal test group.

With these models and BOSS-like spectra, we perform a grid search across our hyperparameter space and find an ideal, global minimum that best standardizes our spectra to look like the original models and recover the original stellar parameters. We perform a similar grid search for a few alternative standardization methods; a constant normalization scheme, standardization by fitting the spectrum with a high order polynomial and the GISIC method. When we compare all methods, we find that our method provides the best recovery of the original stellar parameters, where we see little systematics and recover parameters equally well for most temperatures, metallicities, reddening, and SNR. For this work, we have made a point to track the bias and precision that can result from a given standardization method. While all methods, including ours, show some level of bias, we find that for some specific regions of parameter space the alternative methods perform significantly worse. For GISIC, we see this in poorer precision for metal-poor stars and worse precision for hot stars. For the polynomial, we see this in more bias for metal-rich stars and cool stars.

Next, we use a well-curated list of 100 pc M dwarfs that have a high purity of single stars to validate our method. We find that across the HR diagram, the resulting standardized BOSS spectra look like each other with our Alpha Hulling method. The polynomial method produces better standardized spectra in this validation test, where we see that within a bin on the HR diagram the $\chi^2_\nu$ is lower with this method. However, this is not the most important factor to consider. It is more crucial to understand how well you can recover spectra with
similar parameters among all the other spectra with differing parameters. We used our validation set to see how far away in the HR diagram M dwarf spectra differentiate from each other at a $1\sigma$ level of significance. We found that our method performs much better across the HR diagram, meaning spectra standardized with our method can be better distinguished from other M dwarfs of differing parameters.

Finally, we discuss some of the limitations of the method; primarily that the tuned hyperparameters require that the optical spectra cover the same wavelength range and are sampled on the same wavelength grid as the BOSS spectra. We mitigate some of these issues by resampling the DESI spectra onto the BOSS wavelength grid and clipping the flux values of the BOSS spectra such that both DESI and SDSS were over the same wavelength range. These changes allow us to directly compare spectra from two different surveys with our method. We demonstrate that, after accounting for these limitations, M dwarf spectra from DESI better match the spectra of the same star in SDSS only after they have been standardized with our method. This is one application of our code applied to different surveys, but we stress that this does not mean it is optimized for DESI spectra to be used in a generative model. To do so would require re-optimizing the hyperparameters for the higher resolution and shorter wavelength grid. 
Still, this is an important demonstration of how crucial spectral standardization is when comparing spectra across different surveys and, when properly optimized, facilitates future studies to train better generative models to estimate stellar parameters of M dwarfs.

Overall, our Alpha Hulling method can provide standardized spectra that both produce similar 
looking spectra for M dwarfs of similar stellar parameters and, importantly, can be well distinguished from other M dwarfs of differing parameters. We emphasize that this latter point is crucial and will produce more robust results when our standardization method is used with future template matching or parameter estimation via generative models for these M dwarfs. With this in mind, we strongly recommend that this method is used for the standardization of all BOSS spectra of M dwarfs in SDSS-V. These standardized spectra will provide a strong foundation for future analyses of these low-mass stars and better constrain the critical science questions from SDSS-V.

\begin{acknowledgments}
We thank Jos\'e G. Fern\'andez-Trincado for comments on the paper. We also thank the anonymous referee for a helpful report.

D.S. acknowledges support from the Foundation for Research and Technological Innovation Support of the State of Sergipe (FAPITEC/SE) and the National Council for Scientific and Technological Development (CNPq), under grant numbers 794017/2013 and 444372/2024-5.

B.R-A acknowledges funding support from the ANID Basal project FB210003.

Funding for the Sloan Digital Sky Survey V has been provided by the Alfred P. Sloan Foundation, the Heising-Simons Foundation, the National Science Foundation, and the Participating Institutions. SDSS acknowledges support and resources from the Center for High-Performance Computing at the University of Utah. SDSS telescopes are located at Apache Point Observatory, funded by the Astrophysical Research Consortium and operated by New Mexico State University, and at Las Campanas Observatory, operated by the Carnegie Institution for Science. The SDSS website is \url{www.sdss.org}.

SDSS is managed by the Astrophysical Research Consortium for the Participating Institutions of the SDSS Collaboration, including Caltech, The Carnegie Institution for Science, Chilean National Time Allocation Committee (CNTAC) ratified researchers, The Flatiron Institute, the Gotham Participation Group, Harvard University, Heidelberg University, The Johns Hopkins University, L’Ecole polytechnique f\'{e}d\'{e}rale de Lausanne (EPFL), Leibniz-Institut f\"{u}r Astrophysik Potsdam (AIP), Max-Planck-Institut f\"{u}r Astronomie (MPIA Heidelberg), Max-Planck-Institut f\"{u}r Extraterrestrische Physik (MPE), Nanjing University, National Astronomical Observatories of China (NAOC), New Mexico State University, The Ohio State University, Pennsylvania State University, Smithsonian Astrophysical Observatory, Space Telescope Science Institute (STScI), the Stellar Astrophysics Participation Group, Universidad Nacional Aut\'{o}noma de M\'{e}xico, University of Arizona, University of Colorado Boulder, University of Illinois at Urbana-Champaign, University of Toronto, University of Utah, University of Virginia, Yale University, and Yunnan University.
\end{acknowledgments}

\begin{contribution}
Authors 1--2 (I.M.~and Z.W.) led the analysis, writing, and interpretation. Authors 3--4 (B.R.A.~and G.S.~) provided suggestions and comments on analyses presented in the paper. Author 5 (C.S.) produced the analysis for the correction of offset targets shown in Figure \ref{fig:ex_var_mdwarf}. All other authors (K.G.S.~through A.K.S.) provided thoughtful comments on the content and structure of the paper as it progressed, along with contributing relevant scientific expertise to the project.
\end{contribution}

\software{Astropy \citep{2013A&A...558A..33A,2018AJ....156..123A, astropy:2022}, Numpy \citep{numpy}, Matplotlib \citep{matplotlib}, Scipy \citep{scipy}}

\bibliography{mdwarf_standardization}{}
\bibliographystyle{aasjournal}

\appendix

\section{Alpha Hull Standardization Example}\label{app:example}

We have made our code available for the Alpha Hulling Standardization method described here at \url{https://github.com/imedan/mdwarf_contin/tree/1.0.0}. The default parameters of the standardization class are the optimal hyperparameters found in this work. Below is a short example on how to standardize a BOSS M dwarf spectrum with our code.
\begin{lstlisting}[language=Python]
from astropy.io import fits
from mdwarf_contin.normalize import ContinuumNormalize

# open SDSS-V BOSS spectrum
spec = fits.open('spec-104193-60087-27021598685558821.fits')[1].data

# normalize spectrum
norm = ContinuumNormalize(spec.LOGLAM[spec.LOGLAM > 3.6], spec.FLUX[spec.LOGLAM > 3.6])
norm.find_continuum()

# get the pseudo-continuum in absolute units
continuum = norm.continuum

# standardize the spectrum
stand_spec = norm.flux / continuum
\end{lstlisting}
A fully worked example with plots can also be found in our Github repository at \url{https://github.com/imedan/mdwarf_contin/blob/1.0.0/tests/example_usage.ipynb}.

\end{document}